\PassOptionsToPackage{unicode}{hyperref}
\PassOptionsToPackage{hyphens}{url}
\PassOptionsToPackage{dvipsnames,svgnames,x11names}{xcolor}
\documentclass[
]{article}
\usepackage{amsmath,amssymb}
\usepackage{iftex}
\ifPDFTeX
  \usepackage[T1]{fontenc}
  \usepackage[utf8]{inputenc}
  \usepackage{textcomp} 
\else 
  \usepackage{unicode-math} 
  \defaultfontfeatures{Scale=MatchLowercase}
  \defaultfontfeatures[\rmfamily]{Ligatures=TeX,Scale=1}
\fi
\usepackage{lmodern}
\ifPDFTeX\else
\fi
\IfFileExists{upquote.sty}{\usepackage{upquote}}{}
\IfFileExists{microtype.sty}{
  \usepackage[]{microtype}
  \UseMicrotypeSet[protrusion]{basicmath} 
}{}
\makeatletter
\@ifundefined{KOMAClassName}{
  \IfFileExists{parskip.sty}{%
    \usepackage{parskip}
  }{
    \setlength{\parindent}{0pt}
    \setlength{\parskip}{6pt plus 2pt minus 1pt}}
}{
  \KOMAoptions{parskip=half}}
\makeatother
\usepackage{xcolor}
\usepackage[margin=1in]{geometry}
\usepackage{longtable,booktabs,array}
\usepackage{calc} 
\usepackage{etoolbox}
\makeatletter
\patchcmd\longtable{\par}{\if@noskipsec\mbox{}\fi\par}{}{}
\makeatother
\IfFileExists{footnotehyper.sty}{\usepackage{footnotehyper}}{\usepackage{footnote}}
\makesavenoteenv{longtable}
\usepackage{graphicx}
\makeatletter
\def\maxwidth{\ifdim\Gin@nat@width>\linewidth\linewidth\else\Gin@nat@width\fi}
\def\maxheight{\ifdim\Gin@nat@height>\textheight\textheight\else\Gin@nat@height\fi}
\makeatother
\setkeys{Gin}{width=\maxwidth,height=\maxheight,keepaspectratio}
\makeatletter
\def\fps@figure{htbp}
\makeatother
\providecommand{\tightlist}{%
  \setlength{\itemsep}{0pt}\setlength{\parskip}{0pt}}
\usepackage{xcolor}
\usepackage{authblk}
\usepackage{booktabs}
\usepackage{longtable}
\usepackage{array}
\usepackage{multirow}
\usepackage{wrapfig}
\usepackage{float}
\usepackage{colortbl}
\usepackage{pdflscape}
\usepackage{threeparttable}
\usepackage{threeparttablex}
\usepackage[normalem]{ulem}
\usepackage{makecell}
\usepackage{xcolor}
\ifLuaTeX
  \usepackage{selnolig}  
\fi
\usepackage[]{natbib}
\IfFileExists{bookmark.sty}{\usepackage{bookmark}}{\usepackage{hyperref}}
\IfFileExists{xurl.sty}{\usepackage{xurl}}{} 
\hypersetup{
  pdftitle={Using shrinkage methods to estimate treatment effects in overlapping subgroups in randomized clinical trials with a time-to-event endpoint},
  pdfauthor={Marcel Wolbers, Mar Vázquez Rabuñal, Ke Li, Kaspar Rufibach, Daniel Sabanés Bové},
  colorlinks=true,
  linkcolor={blue},
  filecolor={Maroon},
  citecolor={Blue},
  urlcolor={Blue},
  pdfcreator={LaTeX via pandoc}}

\title{Using shrinkage methods to estimate treatment effects in overlapping subgroups in randomized clinical trials with a time-to-event endpoint}
\author[1]{Marcel Wolbers}
\author[2]{Mar Vázquez Rabuñal}
\author[2]{Ke Li}
\author[1]{Kaspar Rufibach}
\author[3]{Daniel Sabanés Bové}

\affil[1]{Methods,  Collaboration,  and  Outreach  Group,  Product  Development  Data  Sciences, Pharma Development, Roche, Basel, Switzerland}
\affil[2]{Product  Development  Data  Sciences, Pharma Development, Roche, Basel, Switzerland}
\affil[3]{Data Science Acceleration, Product  Development  Data  Sciences, Pharma Development, Roche, Basel, Switzerland}

\date{July 16, 2024}

\begin{document}
\maketitle
\begin{abstract}
In randomized controlled trials, forest plots are frequently used to investigate the homogeneity of treatment effect estimates in pre-defined subgroups. However, the interpretation of subgroup-specific treatment effect estimates requires great care due to the smaller sample size of subgroups and the large number of investigated subgroups. In the literature, Bayesian shrinkage methods have been proposed to address these issues, but they often focus on disjoint subgroups while subgroups displayed in forest plots are overlapping, i.e., each subject appears in multiple subgroups. In our proposed approach, we first build a flexible Cox model based on all available observations, including categorical covariates that identify the subgroups of interest and their interactions with the treatment group variable. We explore both penalized partial likelihood estimation with a lasso or ridge penalty for treatment-by-covariate interaction terms, and Bayesian estimation with a regularized horseshoe prior. One advantage of the Bayesian approach is the ability to derive credible intervals for shrunken subgroup-specific estimates. In a second step, the Cox model is marginalized to obtain treatment effect estimates for all subgroups. We illustrate these methods using data from a randomized clinical trial in follicular lymphoma and evaluate their properties in a simulation study. In all simulation scenarios, the overall mean-squared error is substantially smaller for penalized and shrinkage estimators compared to the standard subgroup-specific treatment effect estimator but leads to some bias for heterogeneous subgroups. A naive overall sample estimator also outperforms the standard subgroup-specific estimator in terms of the overall mean-squared error for all scenarios except for a scenario with substantial heterogeneity. We recommend that subgroup-specific estimators, which are typically displayed in forest plots, are more routinely complemented by treatment effect estimators based on shrinkage methods. The proposed methods are implemented in the R package \texttt{bonsaiforest}.
\end{abstract}

\hypertarget{introduction}{%
\section{Introduction}\label{introduction}}

Exploratory subgroup analyses play a crucial role in the analysis of randomized clinical trials (RCT). These analyses aim to assess the consistency of treatment effects across pre-defined subgroups. Typically, there are multiple variables used to define these subgroups, including randomization stratification factors, demographics, and known prognostic factors. As a result, researchers often examine 20 or more subgroups. Forest plots, which visually represent treatment effect estimates across subgroups, and treatment-by-subgroup interaction tests are common methods used for exploratory subgroup analyses (\citet{alosh2017tutorial}).

While some of the pre-defined subgrouping variables are known to be prognostic for the outcome, there is usually no prior rationale suggesting that any of these variables are predictive, meaning they are associated with heterogeneity of the treatment effect. In fact, if there is a strong suspicion that one of the subgrouping variables, such as a biomarker related to the mechanism of action of the investigated drug, influences the size of the treatment effect, it should be addressed during the trial design phase. For example, one could consider (adaptive) biomarker-stratified designs, enrichment designs, or biomarker-strategy designs (\citet{freidlin2010}, \citet{nguyen2021}).

RCT are typically powered to detect a clinically relevant treatment effect in the overall population and subgroup analyses are exploratory in nature.
There are several challenges associated with subgroup analyses, including small sample sizes of subgroups leading to low precision of treatment effect estimates, low power of treatment-by-subgroup interaction tests, a large number of investigated subgroups, and the tendency of investigators to focus on the most extreme results across subgroups (\citet{alosh2017tutorial}). This led several authors to encourage skepticism towards most reported subgroup effects and claims that the overall treatment effect of a RCT is often a more reliable estimate of the treatment effect in the various subgroups examined than the observed effects in individual subgroups (\citet{yusuf1991jama}, \citet{sleight2000debate}). Published guidance to assess the credibility of subgroup analyses therefore stresses the importance of additional criteria including strong biological plausibility for a differential effect of treatment in the subgroup, pre-specification of the direction of effect modification, and consideration of a small number of effect modifiers only (\citet{sun2014use}, \citet{emaSubgroups2019}, \citet{schandelmaier2020iceman}).

There is an extensive literature on identification of subgroups with an enhanced or diminished treatment effect and the estimation of the treatment effect in those subgroups, see \citet{lipkovich2017tutorial} and \citet{loh2019} for reviews. For treatment effect estimation in pre-defined subgroups specifically, shrinkage estimators have been advocated as a complement to conventional methods (\citet{jones2011}, \citet{henderson2016}, \citet{pennello2018}, \citet{iche17}).
Rather than using data from a specific subgroup only to obtain a treatment effect estimator, shrinkage estimators borrow information across subgroups in order to obtain a more favorable bias-variance trade-off, i.e.~reducing variance substantially at the expense of some bias. The literature on shrinkage estimators has mostly focused on methods for disjoint subgroups. However, subgroups displayed in forest plots, e.g.~those defined by stratification factors, demographics, or prognostic factors, respectively, often overlap, with each subject appearing in multiple subgroups.

In this article we propose shrinkage estimators for overlapping subgroups in RCT with a time-to-event endpoint. We introduce the methodology in Section 2, apply it to a case study in Section 3, and evaluate it in a simulation study in Section 4. We conclude the manuscript with a discussion in Section 5.

\hypertarget{shrinkage-methods-for-treatment-effect-estimation-in-subgroups}{%
\section{Shrinkage methods for treatment effect estimation in subgroups}\label{shrinkage-methods-for-treatment-effect-estimation-in-subgroups}}

We discuss methods applicable to data from a RCT of an intervention versus a control treatment with a time-to-event endpoint and the hazard ratio as the treatment effect measure of interest. The proposed estimators are derived in four steps: First, we define a global outcome model which includes subgroup-by-treatment interactions. Second, we apply a Bayesian shrinkage prior or a lasso/ridge penalty to the interaction terms. The model is then fit using all observations in our RCT.
Third, we standardize the global model to obtain predicted survival curves for both treatment arms in each subgroup. Fourth, we derive an average hazard ratio for each subgroup from these survival curves.

We denote the subjects included in the trial by the subscript \(i\) (\(i=1,\ldots,N\)) and their treatment assignment by the indicator variable \(z_i\) taking values of 0 for control and 1 for intervention. Assume that the subgroups are defined by the levels of \(p\) categorical subgrouping variables \(x_j\) \((j=1,\ldots,p)\), and that subgrouping variable \(x_j\) has \(l_j\) levels. The total number of investigated subgroups is then \(K=\sum_{j=1}^p l_j\). We denote indicator variables for each subgroup by \(s_{ik}\) (i.e.~\(s_{ik}=1\) if subject \(i\) belongs to subgroup \({\cal S}_k\) and 0 otherwise) for \(k=1,\ldots, K\).

\hypertarget{global-outcome-model}{%
\subsection{Global outcome model}\label{global-outcome-model}}

The global outcome model is a Cox proportional hazards model with treatment and all subgrouping variables as main effects, and subgroup-by-treatment interactions. That is, we model the hazard function \(h_i(t)\) of subject \(i\) as

\[
  h_i(t) =  h_0(t) \cdot \exp(\underbrace{\beta_{0}z_i}_{\substack{\text{main treatment} \\ \text{effect}}} 
    + \underbrace{\alpha_1 s_{i1}+\ldots + \alpha_K s_{iK}}_{\substack{\text{main subgroup} \\ \text{effects}}}
    + \underbrace{\beta_1 s_{i1} z_i +\ldots + \beta_K s_{iK}z_i}_{\substack{\text{subgroup-by-treatment} \\ \text{interactions}}})  
\]

Note that this model uses a binary indicator for each level of each subgrouping variable and is therefore overparameterized. The advantage of this parameterization is that all subgroups are treated symmetrically. We did not include higher-order interactions in our global model because our interest focuses on subgroups defined by a single subgrouping variable at a time, as typically displayed in forest plots. As shown in our simulations for the misspecified scenario, treatment effect estimation of our methods in these subgroups does not appear to be adversely impacted if the true data-generating model includes a triple interaction.

We explored both a penalized frequentist and a Bayesian framework for parameter estimation. Frequentist estimation is performed using penalized maximum partial likelihood inference. For estimation using the Bayesian method, the chosen priors for the main treatment and subgroup effects, i.e.~the regression coefficients \(\beta_0\), \(\alpha_1,\ldots,\alpha_K\) are independent normal priors with mean 0 and standard deviation 5, i.e.~essentially flat priors across plausible effect sizes. The baseline hazard was parametrized via M-splines, i.e.~non-negative spline functions, with knots at the quartiles of the observed event times and boundary knots at the minimum and maximum event times (minus or plus the smallest observed difference between event times, respectively). A Dirichlet prior was used for the M-spline coefficients (\citet{brilleman2020}). Penalization and priors for the subgroup-by-treatment interaction terms are described in the next section.

\hypertarget{shrinkage}{%
\subsection{Shrinkage}\label{shrinkage}}

Our methods are applicable to settings without an a priori rationale for effect modification by any of the subgrouping variables. Consequently, we assume exchangeability for all subgroup-by-treatment interaction coefficients \(\beta_1,\ldots,\beta_K\) and further anticipate that in most settings, none or only few of the subgrouping variables are predictive, i.e.~that most of the interaction coefficients are small or even zero.\\
For Bayesian parameter estimation, we chose the regularized horseshoe prior for the interaction coefficients \(\beta_1,\ldots,\beta_K\) (\citet{carvalho2009}, \citet{piironen2017}, \citet{vanErp2019}). The horseshoe prior is a popular shrinkage prior. As described in \citet{carvalho2009}, it has fat Cauchy-like tails allowing strong signals to remain large (that is, un-shrunk) whereas its infinitely tall spike at the origin encourages sparsity of the estimated parameter vector.
As proposed by \citet{piironen2017}, an additional regularization was applied. The global model was implemented with the R package \texttt{brms} (\citet{bruckner2017}) with the prior \texttt{"horseshoe(1)"} for the interaction terms and special family function \texttt{"cox()"} for the outcome model.

For penalized frequentist parameter estimation, we added a lasso penalty, i.e.~\(\lambda \sum_{k=1}^K |\beta_k|\), or a ridge penalty, i.e.~\(\lambda \sum_{k=1}^K \beta_k^2\), to the partial log-likelihood of the Cox model while leaving all other terms unpenalized. The penalty parameter \(\lambda\) was chosen by cross-validation of the partial log-likelihood as implemented in the R package \texttt{glmnet} (\citet{SimonGlmnetCox2011}).

\hypertarget{standardization}{%
\subsection{Standardization}\label{standardization}}

We used standardization (also referred to as G-computation) to derive marginal survival curves in each subgroup for the intervention and control groups, respectively (\citet{hernan2023}). That is, for each subject \(i\), we used parameter estimates from the global outcome model (fit to all subjects in the trial) to predict survival curves \(\hat{S}_{i,C}(t)\) and \(\hat{S}_{i,I}(t)\) for each subject \(i\) by using the subject's covariate values and assuming that (hypothetically) the subject had been assigned to control or intervention, respectively. The marginal survival curve for subgroup \({\cal S}_k\) (with total sample size \(n_k\)) for the intervention group is then estimated via \(\hat{S}_{k,I}(t) = 1/n_k \cdot \sum_{i: s_{ik}=1} \hat{S}_{i,I}(t)\) and analogously for the control group.

For estimation based on the Bayesian method, standardization was based on posterior draws from the global model parameters (including those parametrizing the baseline hazard). For penalized frequentist estimation, standardization used the point estimates of the regression coefficients and the Breslow estimator of the baseline hazard.

\hypertarget{derivation-of-the-average-hazard-ratio-ahr-for-each-subgroup}{%
\subsection{Derivation of the average hazard ratio (AHR) for each subgroup}\label{derivation-of-the-average-hazard-ratio-ahr-for-each-subgroup}}

The hazard functions corresponding to the marginal survival functions in each subgroup for the intervention and control groups may not be proportional (even if the global Cox model is correctly specified) because of non-collapsibility of the hazard ratio (\citet{daniel2021}). To account for this, we used average hazard ratios (AHR) to quantify the treatment effect in subgroups (\citet{kalbfleisch1981}, \citet{schemper2009}, \citet{rauch2018}). For subgroup \({\cal S}_k\), the AHR is defined as
\[ \mathrm{AHR}_k(\gamma, L) = \frac{\int_0^L \hat{S}^{\gamma}_{k,C}(t) d(\hat{S}^{\gamma}_{k,I}(t))}{\int_0^L \hat{S}^{\gamma}_{k,I}(t) d(\hat{S}^{\gamma}_{k,C}(t))}.\]
We selected the power \(\gamma=1\) because for this choice, the AHR corresponds to the odds of concordance, i.e.~it is an interpretable quantity (\citet{schemper2009}).

For Bayesian estimation, a smooth baseline hazard is fit and the numerator of the AHR formula (with \(\gamma=1\)) can be calculated as
\(\int_0^L \hat{S}_{k,C}(t) d(\hat{S}_{k,I}(t)) = \int_0^L \hat{S}_{k,C}(t) \hat{h}_{k,I}(t) \hat{S}_{k,I}(t) dt\) where \(\hat{h}_{k,I}(t)\) is the hazard function corresponding to \(\hat{S}_{k,I}(t)\). The denominator can be calculated analogously.
We chose \(L\) as the largest uncensored event time, and approximated the integrals in the numerator and the denominator of the AHR formula numerically based on Riemann sums using a grid of function values between 0 and L. Based on the posterior draws of the AHR, the subgroup treatment effect estimate and a corresponding 95\% credible interval was then selected as their median, 2.5\% and 97.5\% quantiles.

For the penalized frequentist estimation, the predicted survival curves are right-continuous step-functions with steps at the observed event times \(t_s\), and the integrals become sums over unique event times. For example, the numerator of the AHR formula (with \(\gamma=1\)) can be calculated as \(\int \hat{S}_{k,C}(t) d(\hat{S}_{k,I}(t)) = \sum_{t_s} \hat{S}_{k,C}(t_s) (\hat{S}_{k,I}(t_{s+1})-\hat{S}_{k,I}(t_s))\). The penalized frequentist models provide point estimates of the treatment effect but it is not straightforward to derive corresponding confidence intervals. Even though post-selection inference for the lasso has been an active area of statistical research (\citet{taylor2018}), post-selection inference for regression model coefficients of the global Cox model can still not be easily propagated to inference for the AHR.

\hypertarget{extensions-of-the-methodology-to-other-outcomes}{%
\subsection{Extensions of the methodology to other outcomes}\label{extensions-of-the-methodology-to-other-outcomes}}

It is straightforward to extend the proposed shrinkage estimators to continuous or binary endpoints. Indeed, for these endpoints, the standardization step (which gives standardized means or proportions) is easier and the effect measure of interest (e.g.~a mean difference, an absolute risk difference, or an odds ratio) can be obtained directly.

Binary outcomes are also supported by our software implementation and further discussed in the supplementary materials. In particular, we demonstrate that in this setting, the standard subgroup-specific estimator (without shrinkage) and an estimator based on the global model without penalization of subgroup-by-treatment interactions are mathematically equivalent if a slight modification to the standardization step is applied. In the time-to-event setting this equivalence no longer holds mathematically but, typically, standard subgroup-specific estimators will be numerically similar to estimates based on a global model without penalization.

\hypertarget{case-study-the-gallium-trial}{%
\section{Case study: the GALLIUM trial}\label{case-study-the-gallium-trial}}

The GALLIUM trial randomized 1202 untreated subjects with follicular lymphoma to obinutuzumab-based
chemotherapy versus rituximab-based chemotherapy (\citet{marcus2017}). The primary endpoint was progression-free survival (PFS) as assessed by the investigator and defined as the time from randomization to disease progression or death. Subjects without an event were censored at their last tumor assessment. Due to regional heterogeneity in standard of care, trial sites had to select one of three chemotherapy backbone therapies. Randomization was stratified by the chemotherapy backbone, the Follicular Lymphoma International Prognostic Index (FLIPI) risk group, and the geographic region.

The GALLIUM trial was fully analyzed at a pre-planned efficacy interim analyses after 245 PFS events had been observed. The results showed a significantly prolonged PFS in the obinutuzumab group (hazard ratio 0.66, 95\% CI 0.51-0.85, \(p\)= 0.001). Pre-planned subgroup analyses according to baseline characteristics and stratification factors showed no strong evidence of heterogeneity of treatment effects. However, in the subgroups defined by FLIPI, the observed hazard ratio (95\% CI) was 1.17 (0.63-2.19) in the low, 0.59 (0.37-0.92) in the intermediate, and 0.58 (0.41-0.84) in the high risk category (FLIPI-by-treatment interaction test: \(p=0.14\)) (\citet{marcus2017}). Of note, there had been no a priori clinical rationale for treatment modification by the FLIPI score and the number of PFS events in the FLIPI was relatively low (40 events).

\begin{figure}
\centering
\includegraphics{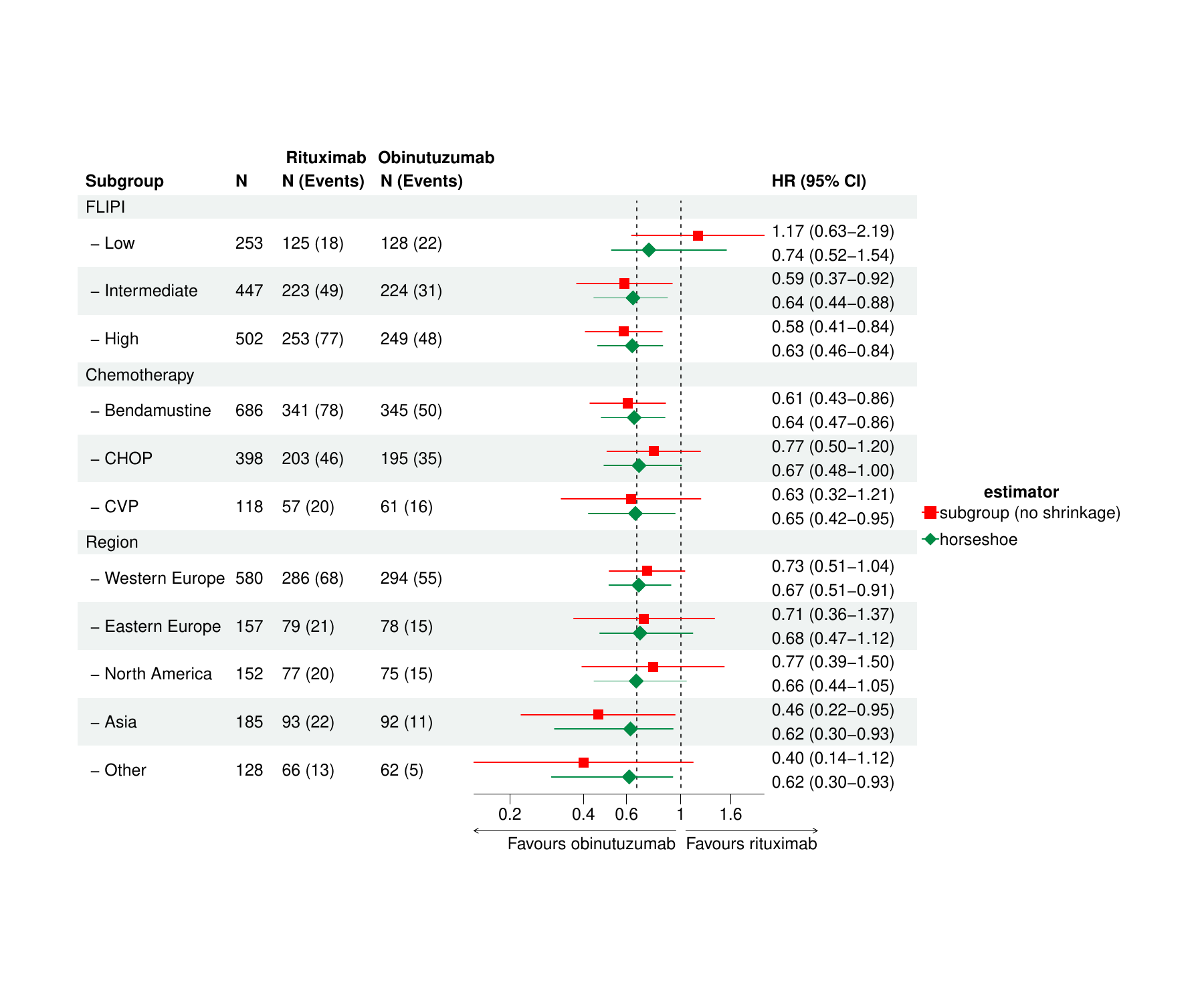}
\caption{\label{fig:unnamed-chunk-2}\label{fig:gallium} Treatment effect estimates in subgroups for the GALLIUM trial (primary analysis data snapshot).}
\end{figure}

A forest plot which complements the standard subgroup-specific treatment effect estimates with shrinkage estimates using the horseshoe prior according to the stratification factors is shown in Figure \ref{fig:gallium}. In the subgroups defined by FLIPI, the horseshoe prior gave average hazard ratios (95\% creditibility intervals) of 0.74 (0.52-1.54) in the low, 0.64 (0.44-0.88) in the intermediate, and 0.63 (0.46-0.84) in the high risk groups. As demonstrated in the simulation study reported in the next section, the shrinkage estimates are expected to have a substantially reduced overall mean-squared error at the expense of some bias.

The final GALLIUM analysis was conducted 10 years after enrollment of the first subject and included 450 PFS event (\citet{townsend2023}). Standard and shrinkage treatment effect estimates in subgroups defined by the stratification factors are reported in the supplementary materials.

\hypertarget{simulation-study}{%
\section{Simulation study}\label{simulation-study}}

We conducted a simulation study to compare the proposed subgroup treatment effect estimators. The simulation parameters were motivated by the GALLIUM trial. We explored six simulation scenarios: one with homogeneous treatment effects across all subgroups and five with treatment effect heterogeneity according to one or several subgrouping variables, respectively.

\hypertarget{candidate-estimators}{%
\subsection{Candidate estimators}\label{candidate-estimators}}

We compared the shrinkage estimators introduced in Section 2, i.e.~the lasso, ridge, and horseshoe estimators, with conventionally used subgroup-specific and population estimators. In addition, the estimators were compared to a published approach for subgroup treatment effect estimation based on Bayesian model averaging.

The standard subgroup-specific and population estimators cover the extremes of no shrinkage versus full shrinkage, respectively. They are included for comparison purposes because they are typically reported in publications of clinical trials. The standard subgroup-specific treatment effect estimator for subgroup \({\cal S}_k\) was obtained by fitting a Cox proportional hazards model with treatment as the only covariate to subjects from subgroup \({\cal S}_k\) only.
This estimator is frequently displayed in forest plots (sometimes with adjustment for prognostic factors not associated with the selected subgroup). The population estimator for subgroup \({\cal S}_k\) was obtained by fitting a Cox proportional hazards model with treatment as the only covariate to all subjects in the trial. The population estimator is by construction identical for all subgroups.

Bayesian model averaging for treatment effect estimation in subgroups was proposed and implemented in the R package \texttt{subtee} by \citet{bornkamp2017} and \citet{ballarini2021}. As described in \citet{bornkamp2017}, this approach also results in shrinkage of treatment effects in subgroups towards the overall treatment effect. In brief, the approach involves fitting \(K\) candidate proportional hazards models \(M_p\) (\(p=1,\ldots, K\)) with a single subgroup-by-treatment interaction:
\[
  M_p: h_i(t) =  h_0(t) \cdot \exp(\beta_{0}z_i + \alpha_1 s_{i1}+\ldots + \alpha_K s_{iK} + \beta_p s_{ip} z_i)
\]
The authors define the treatment effect implied by \(M_p\) for subgroup \({\cal S}_k\) as \(\beta_0 + w \beta_p\) with \(w= |{\cal S}_k\cap {\cal S}_p|/|{\cal S}_p|\), i.e.~they use a simpler approach than our standardization approach which ignores the non-collapsibility of the hazard ratio. The overall posterior distribution of the treatment effect for subgroup \({\cal S}_k\) is then a mixture of the posterior distributions under each model \(M_p\). We chose equal a priori weights for all candidate models (the default in \texttt{subtee}) which are updated using the Bayes information criterion (BIC) approximation for the posterior weights in the mixture distribution.

\hypertarget{data-generation-and-determination-of-the-simulation-ground-truth}{%
\subsection{Data generation and determination of the simulation ground truth}\label{data-generation-and-determination-of-the-simulation-ground-truth}}

We simulated subgrouping variables and time-to-event outcomes for 1:1 randomized trials with an overall sample size of \(n=1'000\) and \(n_{ev}=247\) outcome events included in the analysis. This event number provides 80\% power to detect a true hazard ratio of 0.70 at the two-sided 5\% significance level. The details and statistical models for data generation are provided in the supplementary materials. Here we provide a summary.

Ten subgrouping variables (6 with two levels, 3 with 3 levels, and 1 with 4 levels) defining a total of 25 subgroups of interest were simulated. The subgrouping variables were simulated by categorizing a ten-dimensional multivariate normal distribution with between-variable correlations varying from 0 to 0.5. The sizes of the subgroups ranged from 15\% to 80\% of the total sample size.

The time-to-event outcomes were simulated according to the global outcome model. Specifically, we used a Weibull proportional hazards model.
We assumed that subjects were recruited uniformly over three years, an independent exponential study drop-out distribution with an annual drop-out probability of 2\%, and administrative censoring after the targeted number of events had been observed.

Two subgrouping variables were simulated as being prognostic (with conditional hazard ratios of 0.7 and 1.5, respectively). The overall and subgroup treatment effects were varied across 6 scenarios:

\begin{enumerate}
\def\labelenumi{\arabic{enumi}.}
\tightlist
\item
  \emph{Positive (homogeneous)}: Model without subgroup-by-treatment interactions. AHR in all subgroups are approximately 0.66.
\item
  \emph{Positive (except 1 subgroup)}: Subgroup-by-treatment interaction for one subgrouping variable with 3 levels. AHR for the corresponding subgroups are 1.00, 0.53, and 0.53. AHR for all other subgroups are approximately 0.68.
\item
  \emph{Negative (except 1 subgroup)}: Subgroup-by-treatment interaction for one subgrouping variable with 3 levels. AHR for the corresponding subgroups are 0.50, 1.25, and 1.24. AHR for all other subgroups are approximately 0.98.
\item
  \emph{Mild heterogeneity}: Different AHR in each subgroup with AHR ranging from 0.70 to 1.17.
\item
  \emph{Strong heterogeneity}: Different AHR in each subgroup with AHR ranging from 0.52 to 1.38.
\item
  \emph{Misspecified}: The true data-generating model includes an additional triple interaction between the first 2 (binary) subgrouping variables and treatment which imply AHR between 0.33 and 0.99 in the 4 ``interaction subgroups'' defined by combinations of levels of these two subgrouping variables. The resulting AHR for the 25 subgroups of interest are 0.76 and 0.55 for the levels of the first subgrouping variable, 0.64 and 0.66 for the levels of the second subgrouping variable, and approximately 0.65 for all other subgroups.
\end{enumerate}

The true AHR (reported above) were evaluated using large simulated data sets for each scenario and Kaplan-Meier estimation of the survival curves in each subgroup. Details are provided in the supplementary materials. In the overall population, AHR values ranged between 0.65 and 0.98 across scenarios; the median time-to-event was 4.93 years in the control arm for all scenarios and ranged between 5.08 and 7.27 in the intervention arm.

\hypertarget{evaluation-criteria}{%
\subsection{Evaluation criteria}\label{evaluation-criteria}}

To evaluate the performance of the different estimators, \(n_{sim}=1'000\) simulations were conducted per scenario. For a scenario, we denote the true log(AHR) in subgroup \({\cal S}_k\) by \(\theta_k\) and its estimate based on data from simulation run \(i\) by \(\hat{\theta}_{ik}\). For the overall evaluation of the different estimator, we compared their overall root mean squared error (RMSE), defined as

\[ 
\text{RMSE}_{overall} = \sqrt{\frac{1}{n_{sim}K} \sum_{i=1}^{n_{sim}} \sum_{k=1}^K (\hat{\theta}_{ik}-\theta_k)^2}
\]

for each scenario. In addition, we evaluated RMSE, bias and the frequentist coverage of 95\% confidence intervals (for the standard subgroup estimator) or credible intervals (for the horseshoe estimator) across simulation runs for all subgroups and simulation scenarios separately.

Because the bias of shrinkage methods is expected to be larger for subgroups with a treatment effect which deviates from the population AHR, we considered a subgroup as ``heterogeneous'' if its true log(AHR) deviated by more than \(\log(1.1)\) from the population log(AHR), i.e.~if the subgroup AHR was more than 9.1\% lower or 10\% higher than the population AHR. Otherwise, the subgroup was considered as ``homogeneous''. According to this classification, 29 of the \(6\times 25\) investigated subgroups were heterogeneous (0, 3, 3, 7, 14, and 2 in simulation Scenarios 1-6, respectively; see supplementary Table S2 for details). We then summarized subgroup RMSE, bias, and coverage across all subgroups and scenarios, and for homogeneous and heterogeneous separately.

For Scenario 2 (for which the AHR was 1.00 in one subgroup and \(\leq\) 0.68 for all others), we made two additional explorations: First, for each estimator, we evaluated the probability with which the smallest subgroup treatment effect estimate correctly identified the subgroup in which the intervention did not work. Second, we evaluated how the number of subgrouping variables included in the global model impacts the amount of shrinkage induced by the horseshoe estimator.

\hypertarget{results}{%
\subsection{Results}\label{results}}

\begin{figure}
\centering
\includegraphics{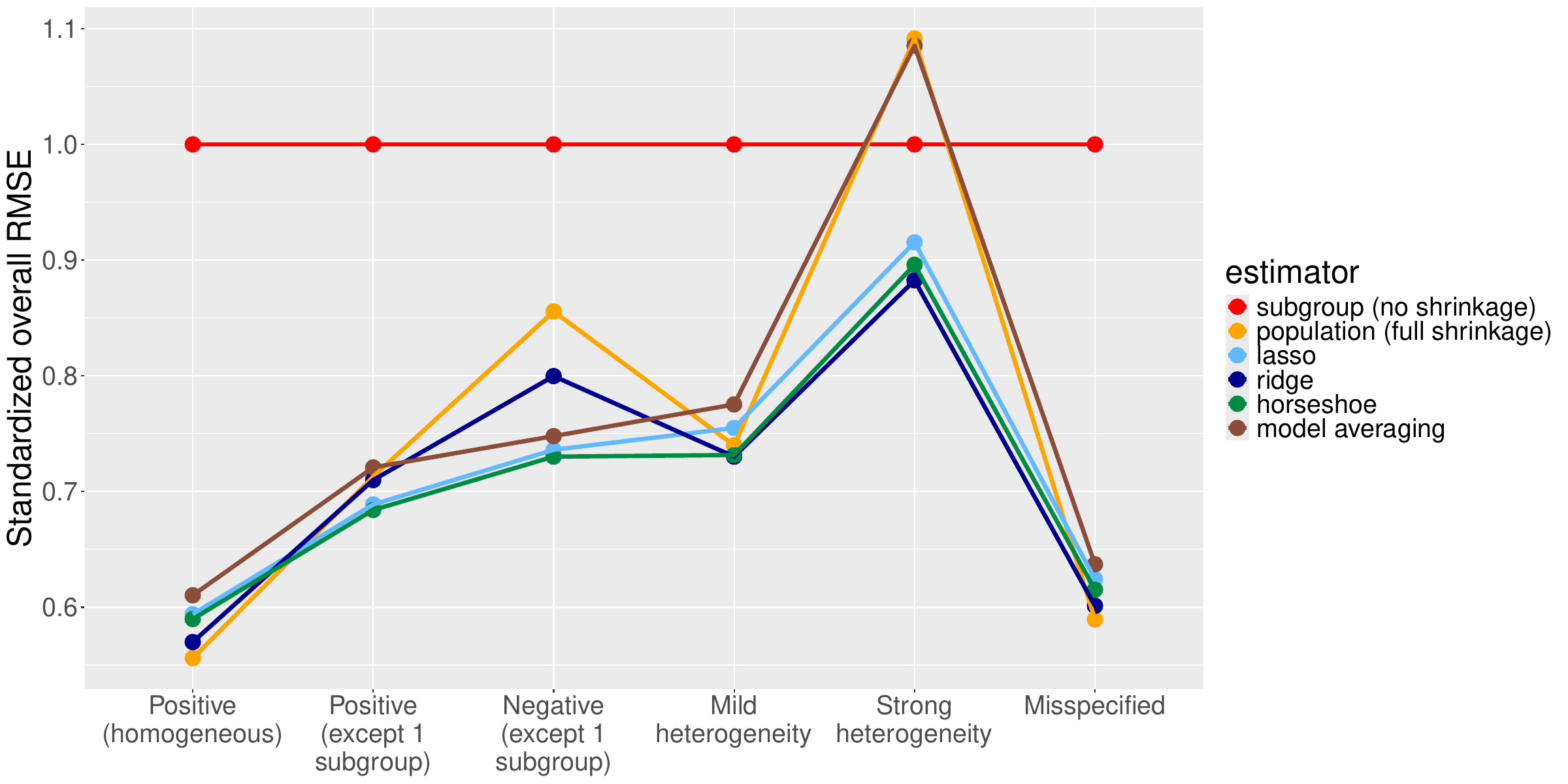}
\caption{\label{fig:unnamed-chunk-4}\label{fig:RMSEoverall} Standardized overall root mean squared error for all estimators of the subgroup log(AHR) across simulation scenarios.}
\end{figure}

The overall RMSEs across estimators and scenarios (relative to the overall RMSE of the standard subgroup estimator without shrinkage) are displayed in Figure \ref{fig:RMSEoverall}. The lasso, ridge, and horseshoe estimators had a lower overall RMSE than the standard subgroup estimator for all scenarios. Specifically, the overall RMSE of the horseshoe estimator was by 10\% to 41\% lower. Interestingly, the population estimator also outperformed the subgroup estimator without shrinkage according to this criterion for all scenarios except for the scenario with strong heterogeneity. The same was true for the Bayesian model averaging estimator which did not manage to outperform the subgroup estimator without shrinkage for the scenario with strong heterogeneity.

\begin{table}

\caption{\label{tab:simulationSummary}Mean (range) of root mean squared error and bias of estimators of the subgroup log(AHR) and observed coverage of confidence or credible interval across all subgroups and simulation scenarios.}
\centering
\begin{tabular}[t]{llll}
\toprule
Criterion & All subgroups & Homogeneous subgroups & Heterogeneous subgroups\\
\midrule
Root mean squared error (RMSE) &  &  & \\
- subgroup (no shrinkage) & 0.23 (0.14-0.36) & 0.22 (0.14-0.36) & 0.24 (0.15-0.34)\\
- population (full shrinkage) & 0.16 (0.13-0.68) & 0.13 (0.13-0.16) & 0.29 (0.17-0.68)\\
- lasso & 0.16 (0.13-0.43) & 0.14 (0.13-0.19) & 0.24 (0.16-0.43)\\
- ridge & 0.16 (0.13-0.53) & 0.14 (0.13-0.20) & 0.24 (0.16-0.53)\\
- horseshoe & 0.16 (0.13-0.43) & 0.14 (0.13-0.19) & 0.23 (0.16-0.43)\\
- model averaging & 0.17 (0.13-0.51) & 0.15 (0.13-0.18) & 0.27 (0.17-0.51)\\
\addlinespace
Absolute bias * &  &  & \\
- subgroup (no shrinkage) & 0.01 (0.00-0.02) & 0.01 (0.00-0.02) & 0.01 (0.00-0.02)\\
- population (full shrinkage) & 0.06 (0.00-0.67) & 0.01 (0.00-0.09) & 0.25 (0.11-0.67)\\
- lasso & 0.04 (0.00-0.34) & 0.01 (0.00-0.08) & 0.14 (0.05-0.34)\\
- ridge & 0.04 (0.00-0.48) & 0.01 (0.00-0.07) & 0.16 (0.06-0.48)\\
- horseshoe & 0.03 (0.00-0.27) & 0.01 (0.00-0.07) & 0.13 (0.05-0.27)\\
- model averaging & 0.05 (0.00-0.42) & 0.02 (0.00-0.10) & 0.17 (0.06-0.42)\\
\addlinespace
Coverage of 95\% confidence or credible interval &  &  & \\
- subgroup (no shrinkage) & 95\% (94\%-97\%) & 95\% (94\%-97\%) & 95\% (94\%-97\%)\\
- horseshoe & 95\% (78\%-99\%) & 96\% (93\%-99\%) & 89\% (78\%-95\%)\\
- model averaging & 92\% (35\%-97\%) & 95\% (90\%-97\%) & 77\% (35\%-93\%)\\
\bottomrule
\multicolumn{4}{l}{\rule{0pt}{1em}* On average, the lasso, ridge and horsehoe estimators reduced the bias across heterogeneneous subgroups by}\\
\multicolumn{4}{l}{\rule{0pt}{1em}     42\%, 35\%, and 47\% compared to the population estimator.}\\
\multicolumn{4}{l}{\rule{0pt}{1em}Summary is across 150 subgroups (25 subgroups x 6 scenarios) including 121 homogeneous and 29 heterogeneous}\\
\multicolumn{4}{l}{\rule{0pt}{1em}subgroups. The Monte Carlo standard error of the estimated coverage is estimated to be 0.7\% given 1'000 simulations}\\
\multicolumn{4}{l}{\rule{0pt}{1em}and true coverage of 95\%.}\\
\end{tabular}
\label{tab:simResults}
\end{table}

RMSE, bias and coverage are summarized in Table \ref{tab:simResults} across all subgroups and simulation scenarios, and separately for homogeneous and heterogeneous subgroups. The average RMSE across all subgroups was largest for the standard subgroup estimator without shrinkage and this was even more pronounced across the 121 homogeneous subgroups. The average RMSE were more similar for the different estimators across the 29 heterogeneous subgroups, with the population estimator being worst and the horseshoe estimator being best. The bias was negligible for the standard subgroup estimator but larger for the other estimators across heterogeneous subgroups. The shrinkage estimators reduced the bias substantially compared to the population estimator. The standard subgroup estimator had adequate coverage for all subgroups. The credible intervals from the horseshoe estimator tended to have higher than nominal frequentist coverage for homogeneous subgroups but reduced coverage for some heterogeneous subgroups (the worst observed coverage among subgroups and simulation scenarios was 78\%). Frequentist coverage of confidence intervals for the Bayesian model averaging estimator was substantially worse than for the horseshoe estimator, especially for the heterogeneous scenarios (supplementary Figure S4). More granular displays of RMSE, bias, and coverage across subgroups and simulation scenarios are shown in the supplementary materials.

For Scenario 2 which included one subgroup where the treatment did not have an effect (subgroup AHR = 1), the estimated probability of correctly identifying that subgroup was
51\% for the standard subgroup estimator, 52\% for the lasso, 20\% for the ridge, 65\% for the horseshoe, and
66\% for the model averaging estimator.

Supplementary Figure S6 illustrates the impact of adding more subgrouping variables to the global model in Scenario 2. When the subgroup is homogeneous, all the estimators are unbiased, but the standard subgroup estimator has a substantially larger variability compared to the horseshoe estimators. Additionally, as more subgrouping variables associated with homogeneous subgroups are included in the model, the variability of the horseshoe estimator slightly decreases. For the subgroup where the treatment did not have an effect (subgroup AHR = 1), the variability of the horseshoe estimators is similar to the standard subgroup estimator and remains relatively constant with an increasing number of subgrouping variables, but there is an increasing bias present.

\hypertarget{discussion}{%
\section{Discussion}\label{discussion}}

We introduced shrinkage methods to obtain treatment effect estimators in overlapping subgroups as typically shown in forest plots. The methods consist of specifying a global outcome model, shrinking subgroup-by-treatment interactions, followed by a standardization step. The lasso and ridge penalties as well as Bayesian methods using the horseshoe prior were explored. We focused on time-to-event endpoints with the (average) hazard ratio as the population-level summary but the methodology is easily extended to other endpoints. An extension to binary endpoints is discussed in the supplementary materials and also supported by our software implementation in the R package \texttt{bonsaiforest} (\citet{bonsaiforest}). The proposed methodology is applicable in the absence of an a priori rationale for effect modification by any of the subgrouping variables. If there is a priori evidence of treatment effect modification, this should inform the prior distribution for the subgroup-by-treatment interaction terms of our approach.

Our simulation study demonstrated that shrinkage methods substantially improve the overall precision of subgroup estimators. Specifically, the overall root mean-squared error (RMSE) of the horseshoe estimator was by 10\% to 41\% lower than the subgroup-specific estimator without shrinkage across all 6 simulation scenarios. Interestingly, the population estimator also outperformed the subgroup estimator without shrinkage according to this criterion for all scenarios except for a scenario with substantial treatment effect heterogeneity which is arguably unlikely to represent the truth in most applications. This supports the intuition of applied statisticians and clinicians that the overall treatment effect of a RCT is often a more reliable estimate of the treatment effect in the various subgroups examined than the observed effects in individual subgroups (\citet{yusuf1991jama},\citet{sleight2000debate}).

All shrinkage estimators reduced the RMSE dramatically in homogeneous subgroups whereas it was similar or larger than the standard subgroup estimator without shrinkage for heterogeneous subgroups. As expected, shrinkage introduced some bias for heterogeneous subgroups. Compared to the population estimator, the horseshoe estimator reduced bias on average by 47\%. Interestingly, the horseshoe estimator was more likely to identify an inefficacious subgroup compared to the subgroup-specific estimator in simulation scenario 2. Differences between the lasso, ridge, and horseshoe estimators were not dramatic but in most scenarios, the horseshoe estimator performed best. The horseshoe estimator also allows to quantify uncertainty via credible intervals in a straightforward way and is therefore our recommended estimator though it is acknowledged that the frequentist coverage of credible intervals may be substantially below their nominal levels.
The model averaging estimator performed substantially worse than the horseshoe estimator, especially in the heterogeneous scenarios. One possible explanation for this is that the candidate models over which the method averaged included only one binary treatment-by-subgroup interaction term at a time. This might not be optimal for subgrouping variables with more than two levels. It also implies that none of the candidate models represented the true data-generating mechanism for any scenario except for the homogeneous scenario. Moreover, as mentioned in the methods section, the model averaging of treatment effects in the chosen software implementation did not account for the non-collapsibility of the hazard ratio.

Our methodology combines ideas from the existing literature: Many authors have advocated shrinkage methods for the analysis of non-overlapping subgroup (\citet{jones2011}, \citet{henderson2016}, \citet{pennello2018}, \citet{iche17}). Similarly, several authors have proposed global Cox regression models with penalization or shrinkage applied to subgroup-by-treatment interactions (\citet{lipkovich2017tutorial}, \citet{ternes2017identification}, \citet{qi2022}). However, to our knowledge, we are the first to apply standardization to obtain subgroup treatment effect estimators based on such models.

We acknowledge several limitations of our manuscript.
First, we focus on the (average) hazard ratio as the summary measure of interest because this is still the most commonly reported summary measure for time-to-event endpoints in clinical trials. As discussed in the literature, hazard ratios are difficult to interpret causally and may show puzzling behavior in subgroups defined by strong prognostic factors (\citet{liu2022},\citet{didelez2022}). Conceptually, our approach could be generalized to alternative summary measure such as the acceleration factor, a difference in milestone probabilities, or a difference in the restricted mean survival times (\citet{gregson2019}). However, this would require modification of the global outcome model, e.g.~to an accelerated failure time models or to a more flexible hazards model.
Second, a large number of methods have been proposed for subgroup identification and the estimation of the treatment effect in those subgroups (\citet{lipkovich2017tutorial},\citet{loh2019}). We regard our methodology as particularly suited for the setting of forest plots where typically a moderate number of categorical subgroups is considered. Methods which are based on more complex machine learning techniques such as trees or random forests and methods which focus on identifying a particularly efficacious subgroup may be more difficult to justify for this setting. However, we acknowledge that we have not formally compared our methods to these competing approaches.
Third, we have not covered the setting of continuous covariates, treatment effect estimation in subgroups defined by multiple covariates, or the complications induced by missing covariate data.

In conclusion, we recommend that subgroup-specific estimators, which are typically displayed in forest plots, are more routinely complemented by estimators based on shrinkage methods. We have presented one proposal to achieve this and encourage alternative proposals and more comprehensive methods comparison studies.

\hypertarget{software-implementation-and-data-availability}{%
\section*{Software implementation and data availability}\label{software-implementation-and-data-availability}}
\addcontentsline{toc}{section}{Software implementation and data availability}

All estimators have been implemented in the R package \texttt{bonsaiforest} which is available on CRAN (\citet{bonsaiforest}). The simulation code is available in the corresponding github repository: \url{https://github.com/insightsengineering/bonsaiforest/tree/main/simulations}.

Qualified researchers may request access to individual patient level data of the GALLIUM trial (\citet{marcus2017}, \citet{townsend2023}) through the clinical study data request platform \url{https://vivli.org}. Further details on Roche's criteria for eligible studies are available here: \url{https://vivli.org/members/ourmembers}. For further details on Roche's Global Policy on the Sharing of Clinical Information and how to request access to related clinical study documents, see here: \url{https://www.roche.com/research_and_development/who_we_are_how_we_work/clinical_trials/our_commitment_to_data_sharing.htm}.

\hypertarget{appendix-appendix}{%
\appendix}

\setcounter{table}{0}  \renewcommand{\thetable}{S\arabic{table}} \setcounter{figure}{0} \renewcommand{\thefigure}{S\arabic{figure}}

\hypertarget{methods-for-binary-outcomes}{%
\section{Supplementary Materials}\label{supplementary-materials}}

\hypertarget{methods-for-binary-outcomes}{%
\subsection{Methods for binary outcomes}\label{methods-for-binary-outcomes}}

\hypertarget{treatment-effect-estimators-in-subgroups-for-binary-outcomes}{%
\textbf{Treatment effect estimators in subgroups for binary outcomes}\label{treatment-effect-estimators-in-subgroups-for-binary-outcomes}}

The treatment effect estimators in subgroups for survival outcomes (with the hazard ratio as the treatment effect measure) discussed in the main paper can be adapted to the simpler setting of binary outcomes (with the odds ratio, the relative risk, or the absolute risk difference as the treatment effect measure) in a straightforward way. This adaptation proceeds by replacing the Cox model by the logistic model.
For binary outcomes, standardization leads to marginal event probabilities in each subgroup for the intervention and control groups, respectively, from which the treatment effect in each subgroup can be directly calculated.

\hypertarget{relation-between-the-standard-subgroup-specific-treatment-effect-estimator-and-an-estimator-based-on-standardization-of-an-unpenalized-global-model-for-binary-outcomes}{%
\textbf{Relation between the standard subgroup-specific treatment effect estimator and an estimator based on standardization of an unpenalized global model for binary outcomes}\label{relation-between-the-standard-subgroup-specific-treatment-effect-estimator-and-an-estimator-based-on-standardization-of-an-unpenalized-global-model-for-binary-outcomes}}

Using the notation from the main manuscript, consider a global logistic regression model of the form

\[
  P(y_i=1) =  \textrm{expit}(\alpha_0+\underbrace{\beta_{0}z_i}_{\substack{\text{main treatment} \\ \text{effect}}} 
    + \underbrace{\alpha_1 s_{i1}+\ldots + \alpha_K s_{iK}}_{\substack{\text{main subgroup} \\ \text{effects}}}
    + \underbrace{\beta_1 s_{i1} z_i +\ldots + \beta_K s_{iK}z_i}_{\substack{\text{subgroup-by-treatment} \\ \text{interactions}}}) 
\]

where \(z_i\) describes treatment assignment and and \(s_{ik}\) are binary indicator variables for subgroups. Denote the observed outcome of subject \(i\) by \(y_i\) \((i=1,\ldots,N)\) and their predicted outcome probabilities from the logistic model by \(\hat{y}_i\).

For any subset \(S\) of subjects of the form \(\{i: z_i=a \mbox{ and } s_{ik}=1\}\) (with \(a=0\) or \(a=1\)), the predicted event probability in \(S\) obtained via ``standardization'' from the global model is identical to the observed probability, i.e.~\(1/|S| \cdot \sum_{i\in S} \hat{y}_i = 1/|S|\cdot \sum_{i\in S} y_i\). To demonstrate this, note that for the over-parametrized logistic model above, an equivalent properly parametrized logistic model exists which includes an indicator variable \(s_i\) for subjects in \(S\) with corresponding regression coefficient \(\beta_s\). The claim then follows directly from the estimating equations for this logistic model, i.e.~by setting the derivative of the log-likelhood of this logistic model with respect to \(\beta_s\) to 0 (see formula (4.25), page 137, in \citet{agresti2012}).

Note that this does not imply that the estimator based on standardization of a global model for binary outcomes proposed in this manuscript is mathematically identical to the standard subgroup-specific estimator if no penalization is applied to the global model. The reason for this is that ``standardization'' here includes only covariate information from subjects with \(s_{ik} = 1\) from the evaluated treatment group (\(z_i=a\)). In contrast, the proposed subgroup treatment effect estimators exploit covariate information from all subjects in a subgroup regardless of their assigned treatment group.

\hypertarget{subgroup-estimates-for-the-gallium-trial-final-analysis-data-snapshot}{%
\subsection{Subgroup estimates for the Gallium trial (final analysis data snapshot)}\label{subgroup-estimates-for-the-gallium-trial-final-analysis-data-snapshot}}

The final GALLIUM analysis was conducted 10 years after enrollment of the first subject and included 450 PFS event (\citet{townsend2023}). A forest plot which complements the standard subgroup-specific treatment effect estimates with shrinkage estimates using the horseshoe prior according to the stratification factors is shown in Figure S1.

\begin{figure}
\centering
\includegraphics{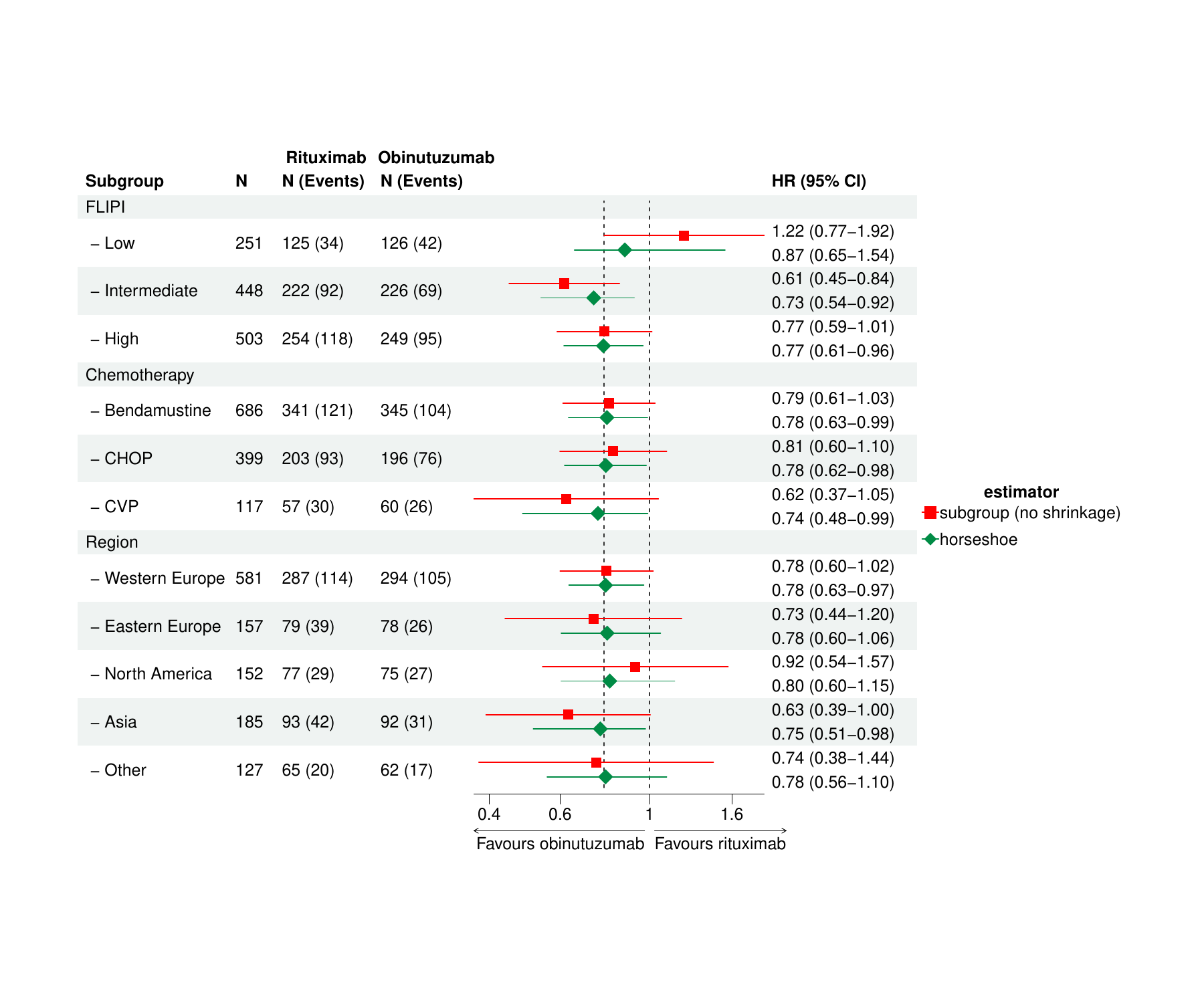}
\caption{\label{fig:unnamed-chunk-2}\label{fig:galliumFinal}Treatment effect estimates in subgroups for the Gallium trial (final analysis data snapshot).}
\end{figure}

\hypertarget{additional-information-on-the-simulation-study-for-time-to-event-outcomes}{%
\subsection{Additional information on the simulation study for time-to-event outcomes}\label{additional-information-on-the-simulation-study-for-time-to-event-outcomes}}

\hypertarget{data-generation}{%
\subsubsection{Data generation}\label{data-generation}}

\hypertarget{sample-size-and-number-of-events}{%
\textbf{Sample size and number of events}\label{sample-size-and-number-of-events}}

We simulated subgrouping variables and time-to-event outcomes for 1:1 randomized trials with total sample size \(n=1000\) with a target number of \(n_{ev}=247\) outcome events included in the analysis.

\hypertarget{simulation-of-subgrouping-variables-and-treatment-arm}{%
\textbf{Simulation of subgrouping variables and treatment arm}\label{simulation-of-subgrouping-variables-and-treatment-arm}}

Subgrouping variables were simulated by categorizing a (latent) ten-dimensional multivariate normal distribution. Continuous outcomes were simulated as

\[ (X_1,\ldots,X_{10})^t \sim \mathcal{N} (\mu, \Sigma)  \]

with mean \(\mu=(0,\ldots,0)^t\). We assumed variances of 1 for all continuous variables, a correlations 0.25 between variables \(X_6-X_8\), a correlation of 0.5 between variables \(X_9\) and \(X_{10}\), and a correlation of 0 for all other associations.

Subgroups were obtained by categorizing latent continuous variables according to appropriate quantiles of the standard normal distribution. The number of categories and their respective size is provided in Table S1.

\begin{table}[h]
\centering
\begin{tabular}{ cccc }
  \hline
  Variable & levels & proportion per level & subgroup indicators \\ \hline
  $X_1$     & 2 & 0.5, 0.5                & $s_{1a}, s_{1b}$ \\
  $X_2$     & 2 & 0.4, 0.6                & $s_{2a}, s_{2b}$ \\
  $X_3$     & 2 & 0.2, 0.8                & $s_{3a}, s_{3b}$ \\
  $X_4$     & 3 & 0.3, 0.3, 0.4           & $s_{4a}, s_{4b}, s_{4c}$ \\
  $X_5$     & 4 & 0.15, 0.15, 0.3, 0.4    & $s_{5a}, s_{5b}, s_{5c}, s_{5d}$ \\
  $X_6$     & 2 & 0.4, 0.6                & $s_{6a}, s_{6b}$ \\
  $X_7$     & 2 & 0.4, 0.6                & $s_{7a}, s_{7b}$ \\
  $X_8$     & 3 & 0.2, 0.3, 0.5           & $s_{8a}, s_{8b}$ \\
  $X_9$     & 2 & 0.2, 0.8                & $s_{9a}, s_{9b}$ \\
  $X_{10}$  & 3 & 0.2, 0.3, 0.5           & $s_{10a}, s_{10b}, s_{10c}$ \\ \hline
\end{tabular}
\label{tab_sgvariables}
\caption{Subgrouping variables in the simulated datasets.}
\end{table}

Subjects were simulated to be assigned alternatively to the control and intervention arm, respectively, giving sample size of 500 per arm. The assigned treatment for subject \(i\) is denoted by \(z_i\) with \(z_i=1\) denoting intervention, and \(z_i=0\) control.

\hypertarget{simulation-of-time-to-event-data-and-censoring}{%
\textbf{Simulation of time-to-event data and censoring}\label{simulation-of-time-to-event-data-and-censoring}}

Uncensored time-to-event outcomes were simulated according to the global outcome model. Specifically, we used the following Weibull accelerated failure time model (which is also a proportional hazards model) to simulate uncensored event times \(T_i\) for subjects \(i\):

\[   \log(T_i) =  \alpha_0 + \beta_0 z_i + \sum_{j=1}^{10}\sum_k \alpha_{jk}s_{jk,i}+ \sum_{j=1}^{10}\sum_k \beta_{jk} s_{jk,i}z_i + \sigma \log(W_i) \mbox{ with } W_i \sim \mathrm{Exponential}(1). \]

For all scenarios, we set \(\alpha_0=2\), \(\sigma=0.85\), and assumed two prognostic factors: \(\alpha_{4c}=-\log(0.7)\sigma\) and \(\alpha_{6b}=-\log(1.5)\sigma\). This implies that the (conditional) hazard ratios associated with level \(c\) of subgrouping variable 4 and level \(b\) of subgrouping variable 6, respectively, are 0.7 and 1.5. The other parameters were varied across 6 scenarios as described below. (All parameters which are not mentioned were set to 0.)

\begin{enumerate}
\def\labelenumi{\arabic{enumi}.}
\tightlist
\item
  \emph{Positive (homogeneous)}: \(\beta_0= -\log(0.66)\sigma\).
\item
  \emph{Positive (except 1 subgroup)}: \(\beta_0= -\log(0.66)\sigma\), \(\beta_{4a}=+\log(0.66)\sigma\), \(\beta_{4b}=-\log(0.8)\sigma\), \(\beta_{4c}=-\log(0.8)\sigma\).
\item
  \emph{Negative (except 1 subgroup)}: \(\beta_0= 0\), \(\beta_{4a}=-\log(0.5)\sigma\), \(\beta_{4b}=-\log(1.25)\sigma\), \(\beta_{4c}=-\log(1.25)\sigma\).
\item
  \emph{Mild heterogeneity}: \(\beta_0= 0\); the regression coefficients corresponding to the 25 treatment-by-subgroup interaction terms were simulated as \(\beta_{jk} = -\log(\sigma)\gamma_{jk}\) with \(\gamma_{jk}\sim \mathcal{N}(0,sd=0.15)\). The same parameters were chosen for all repeated simulations, i.e.~the regression coefficients were not re-drawn for each simulation run.
\item
  \emph{Strong heterogeneity}: Regression parameters were chosen in the same way as for scenario 4 except that the standard deviation of the normal draws was increased from 0.15 to 0.3.
\item
  \emph{Mis-specified}: The data generating model included included an additional triple interaction between the first 2 (binary) subgrouping variables and treatment with corresponding regression coefficients \(\beta_0= -\log(0.66)\sigma\), \(\beta_{1a,2a}= -log(1.5)\sigma\), \(\beta_{1a,2b}= -log(0.92)\sigma\), \(\beta_{1b,2a}= -log(0.5)\sigma\), and \(\beta_{1b,2b}= -log(1.07)\sigma\).
\end{enumerate}

We assumed that subjects were recruited uniformly over a period of three years, an independent exponential study drop-out distribution with an annual drop-out probability of 2\%, and administrative censoring after the targeted number of events had been observed.

\hypertarget{evaluation-of-the-true-subgroup-average-hazard-ratios-ahr-for-each-scenario}{%
\textbf{Evaluation of the true subgroup average hazard ratios (AHR) for each scenario}\label{evaluation-of-the-true-subgroup-average-hazard-ratios-ahr-for-each-scenario}}

To calculate true AHR for each subgroup, we simulated 10 large data sets (\(n=1'000'000\), \(n_{ev}=247'000\)) for each scenario, calculated AHR for each subgroup, and then averaged the corresponding log-transformed AHR across the 10 repetitions. This approach is computationally more efficient than simulating a ten-fold larger data set a single time. The AHR for each subgroup based on these large data sets were estimated by approximating the survival curves by Kaplan-Meier estimates and integrating the resulting integrals from 0 up to the 95\% quantile of the observed event times (to avoid the impact of instable Kaplan-Meier estimators in the tails).

The resulting true AHR for all subgroups and scenarios are shown in Table S2.

\begin{table}
\centering
\caption{\label{tab:trueHR}True AHR in subgroups and the overall population for all scenarios.}
\centering
\begin{tabular}[t]{lllllll}
\toprule
Subgroup & \makecell[l]{Positive \\ (homogeneous)} & \makecell[l]{Positive \\ (except 1 subgroup)} & \makecell[l]{Negative \\ (except 1 subgroup)} & \makecell[l]{Mild \\ heterogeneity} & \makecell[l]{Strong \\ heterogeneity} & \makecell[l]{Mis- \\ specified}\\
\midrule
Overall & $0.66$ & $0.68$ & $0.98$ & $0.88$ & $0.83$ & $0.65$\\
\cellcolor{gray!10}{$S_{1a}$} & \cellcolor{gray!10}{$0.67$} & \cellcolor{gray!10}{$0.68$} & \cellcolor{gray!10}{$0.98$} & \cellcolor{gray!10}{$0.74^*$} & \cellcolor{gray!10}{$0.58^*$} & \cellcolor{gray!10}{$0.76^*$}\\
\cellcolor{gray!10}{$S_{1b}$} & \cellcolor{gray!10}{$0.66$} & \cellcolor{gray!10}{$0.68$} & \cellcolor{gray!10}{$0.98$} & \cellcolor{gray!10}{$1.03^*$} & \cellcolor{gray!10}{$1.10^*$} & \cellcolor{gray!10}{$0.55^*$}\\
$S_{2a}$ & $0.66$ & $0.68$ & $0.98$ & $0.79^*$ & $0.65^*$ & $0.64$\\
$S_{2b}$ & $0.66$ & $0.68$ & $0.98$ & $0.95$ & $0.95^*$ & $0.66$\\
\cellcolor{gray!10}{$S_{3a}$} & \cellcolor{gray!10}{$0.66$} & \cellcolor{gray!10}{$0.68$} & \cellcolor{gray!10}{$0.98$} & \cellcolor{gray!10}{$1.17^*$} & \cellcolor{gray!10}{$1.38^*$} & \cellcolor{gray!10}{$0.65$}\\
\cellcolor{gray!10}{$S_{3b}$} & \cellcolor{gray!10}{$0.66$} & \cellcolor{gray!10}{$0.68$} & \cellcolor{gray!10}{$0.98$} & \cellcolor{gray!10}{$0.82$} & \cellcolor{gray!10}{$0.70^*$} & \cellcolor{gray!10}{$0.65$}\\
$S_{4a}$ & $0.66$ & $1.00^*$ & $0.50^*$ & $0.88$ & $0.82$ & $0.65$\\
$S_{4b}$ & $0.66$ & $0.53^*$ & $1.25^*$ & $0.86$ & $0.78$ & $0.65$\\
$S_{4c}$ & $0.66$ & $0.53^*$ & $1.24^*$ & $0.90$ & $0.87$ & $0.65$\\
\cellcolor{gray!10}{$S_{5a}$} & \cellcolor{gray!10}{$0.67$} & \cellcolor{gray!10}{$0.68$} & \cellcolor{gray!10}{$0.98$} & \cellcolor{gray!10}{$0.96$} & \cellcolor{gray!10}{$0.97^*$} & \cellcolor{gray!10}{$0.65$}\\
\cellcolor{gray!10}{$S_{5b}$} & \cellcolor{gray!10}{$0.67$} & \cellcolor{gray!10}{$0.68$} & \cellcolor{gray!10}{$0.98$} & \cellcolor{gray!10}{$1.12^*$} & \cellcolor{gray!10}{$1.30^*$} & \cellcolor{gray!10}{$0.65$}\\
\cellcolor{gray!10}{$S_{5c}$} & \cellcolor{gray!10}{$0.66$} & \cellcolor{gray!10}{$0.68$} & \cellcolor{gray!10}{$0.98$} & \cellcolor{gray!10}{$0.84$} & \cellcolor{gray!10}{$0.74^*$} & \cellcolor{gray!10}{$0.65$}\\
\cellcolor{gray!10}{$S_{5d}$} & \cellcolor{gray!10}{$0.66$} & \cellcolor{gray!10}{$0.68$} & \cellcolor{gray!10}{$0.98$} & \cellcolor{gray!10}{$0.80$} & \cellcolor{gray!10}{$0.68^*$} & \cellcolor{gray!10}{$0.65$}\\
$S_{6a}$ & $0.67$ & $0.68$ & $0.98$ & $0.95$ & $0.96^*$ & $0.65$\\
$S_{6b}$ & $0.66$ & $0.67$ & $0.98$ & $0.85$ & $0.76$ & $0.65$\\
\cellcolor{gray!10}{$S_{7a}$} & \cellcolor{gray!10}{$0.66$} & \cellcolor{gray!10}{$0.68$} & \cellcolor{gray!10}{$0.98$} & \cellcolor{gray!10}{$0.91$} & \cellcolor{gray!10}{$0.89$} & \cellcolor{gray!10}{$0.66$}\\
\cellcolor{gray!10}{$S_{7b}$} & \cellcolor{gray!10}{$0.66$} & \cellcolor{gray!10}{$0.68$} & \cellcolor{gray!10}{$0.98$} & \cellcolor{gray!10}{$0.86$} & \cellcolor{gray!10}{$0.79$} & \cellcolor{gray!10}{$0.65$}\\
$S_{8a}$ & $0.67$ & $0.68$ & $0.98$ & $0.70^*$ & $0.52^*$ & $0.66$\\
$S_{8b}$ & $0.66$ & $0.68$ & $0.98$ & $0.99^*$ & $1.01^*$ & $0.65$\\
$S_{8c}$ & $0.66$ & $0.68$ & $0.98$ & $0.89$ & $0.84$ & $0.65$\\
\cellcolor{gray!10}{$S_{9a}$} & \cellcolor{gray!10}{$0.66$} & \cellcolor{gray!10}{$0.68$} & \cellcolor{gray!10}{$0.98$} & \cellcolor{gray!10}{$0.88$} & \cellcolor{gray!10}{$0.84$} & \cellcolor{gray!10}{$0.65$}\\
\cellcolor{gray!10}{$S_{9b}$} & \cellcolor{gray!10}{$0.66$} & \cellcolor{gray!10}{$0.68$} & \cellcolor{gray!10}{$0.98$} & \cellcolor{gray!10}{$0.88$} & \cellcolor{gray!10}{$0.82$} & \cellcolor{gray!10}{$0.65$}\\
$S_{10a}$ & $0.66$ & $0.68$ & $0.98$ & $0.96$ & $0.96^*$ & $0.65$\\
$S_{10b}$ & $0.67$ & $0.68$ & $0.98$ & $0.85$ & $0.77$ & $0.65$\\
$S_{10c}$ & $0.66$ & $0.68$ & $0.98$ & $0.87$ & $0.80$ & $0.65$\\
\bottomrule
\multicolumn{7}{l}{\rule{0pt}{1em}Heterogenous subgroup AHR, defined as AHR which differ by more then -9.1\% or +10\% from the population AHR,}\\
\multicolumn{7}{l}{\rule{0pt}{1em}are marked with a star (*).}\\
\end{tabular}
\end{table}

\hypertarget{additional-summaries-of-the-simulation-results-and-other-explorations}{%
\subsubsection{Additional summaries of the simulation results and other explorations}\label{additional-summaries-of-the-simulation-results-and-other-explorations}}

Root mean-squared error, bias, and frequentist coverage of confidence or credible intervals for all estimators stratified by subgroup and scenario are shown in Figures S2-S4. In addition, we visualized the observed RMSE (averaged across subgroups) for each simulation run \(i\), i.e.~

\[ 
\text{RMSE}_{i} = \sqrt{\frac{1}{K} \sum_{k=1}^K (\hat{\theta}_{ik}-\theta_k)^2} 
\]

(using the same notation as in the main manuscript), for all estimators and scenarios via boxplots (Figure S5).

Finally, in an additional exploration for Scenario 2, we assessed how the number of subgrouping variables included in the global model affects the horsehoe estimator in the only subgroup in which the treatment did not work (i.e.~the true AHR was 1.00). To achieve this, we simulated time-to-event outcomes as per Scenario 2 but varied the number of subgroups as follows:

\begin{enumerate}
\def\labelenumi{\arabic{enumi}.}
\tightlist
\item
  Only the single subgrouping variable with 3 level in which heterogeneity is observed is included in the global model (1 subgrouping variable, 3 subgroups).
\item
  The first 4 subgrouping variables are included in the global model (4 subgrouping variables, 9 subgroups).
\item
  Original scenario with 10 subgrouping variables and 25 subgroups.
\item
  Original scenario with an additional 10 subgrouping variables simulated using the same data-generating model as for the first 10 subgrouping variables (20 subgrouping variables, 50 subgroups).
\end{enumerate}

Boxplots of the estimated AHR (across 1'000 simulation runs) for one homogeneous subgroup and for the only subgroup in which the treatment did not work are shown in Figure S6. The more subgroups were included in the global model, the more the horseshoe estimator was shrunken towards the population estimator.

\begin{figure}
\centering
\includegraphics{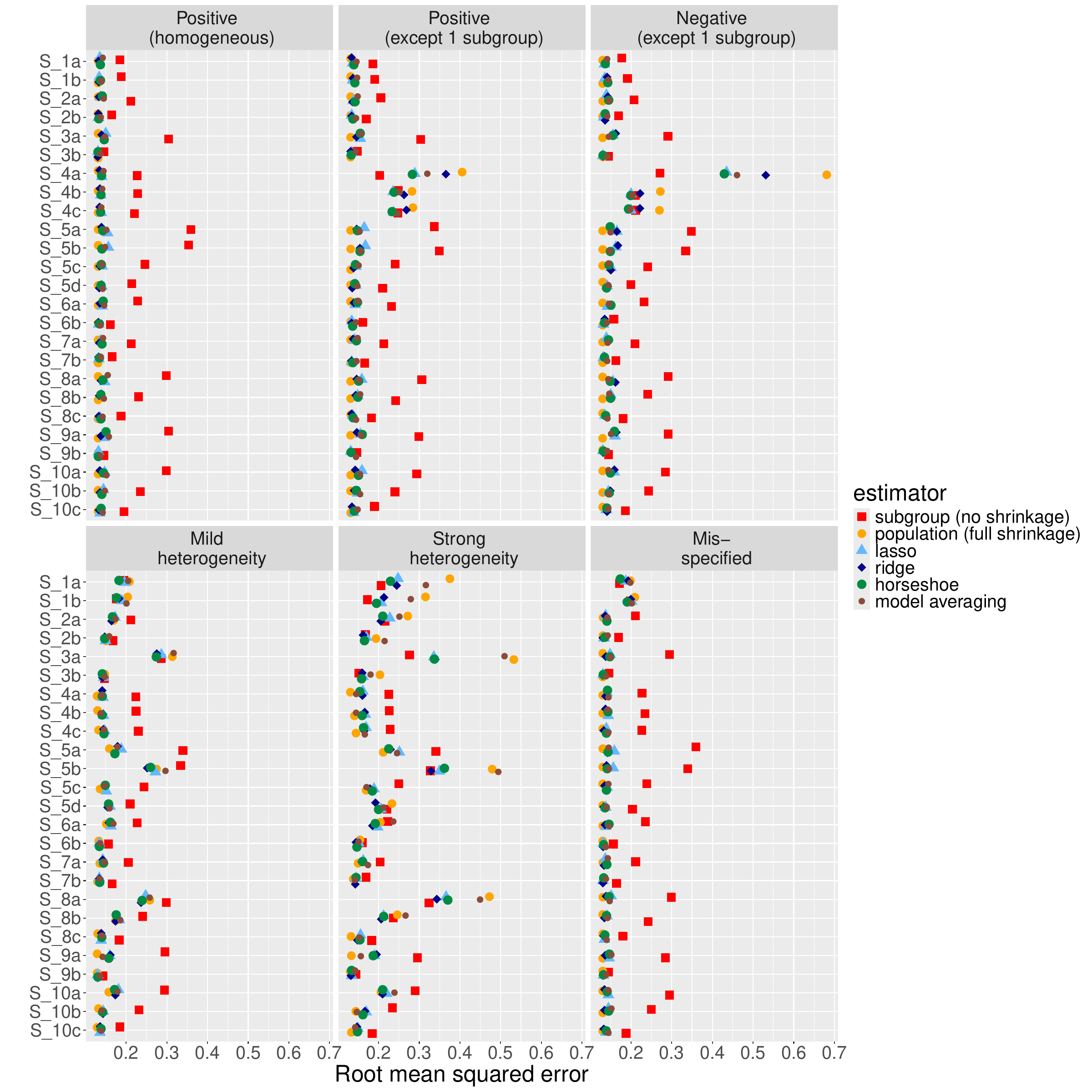}
\caption{\label{fig:unnamed-chunk-3}\label{fig:RMSE}Root mean squared error for all simulation scenarios and subgroups.}
\end{figure}

\begin{figure}
\centering
\includegraphics{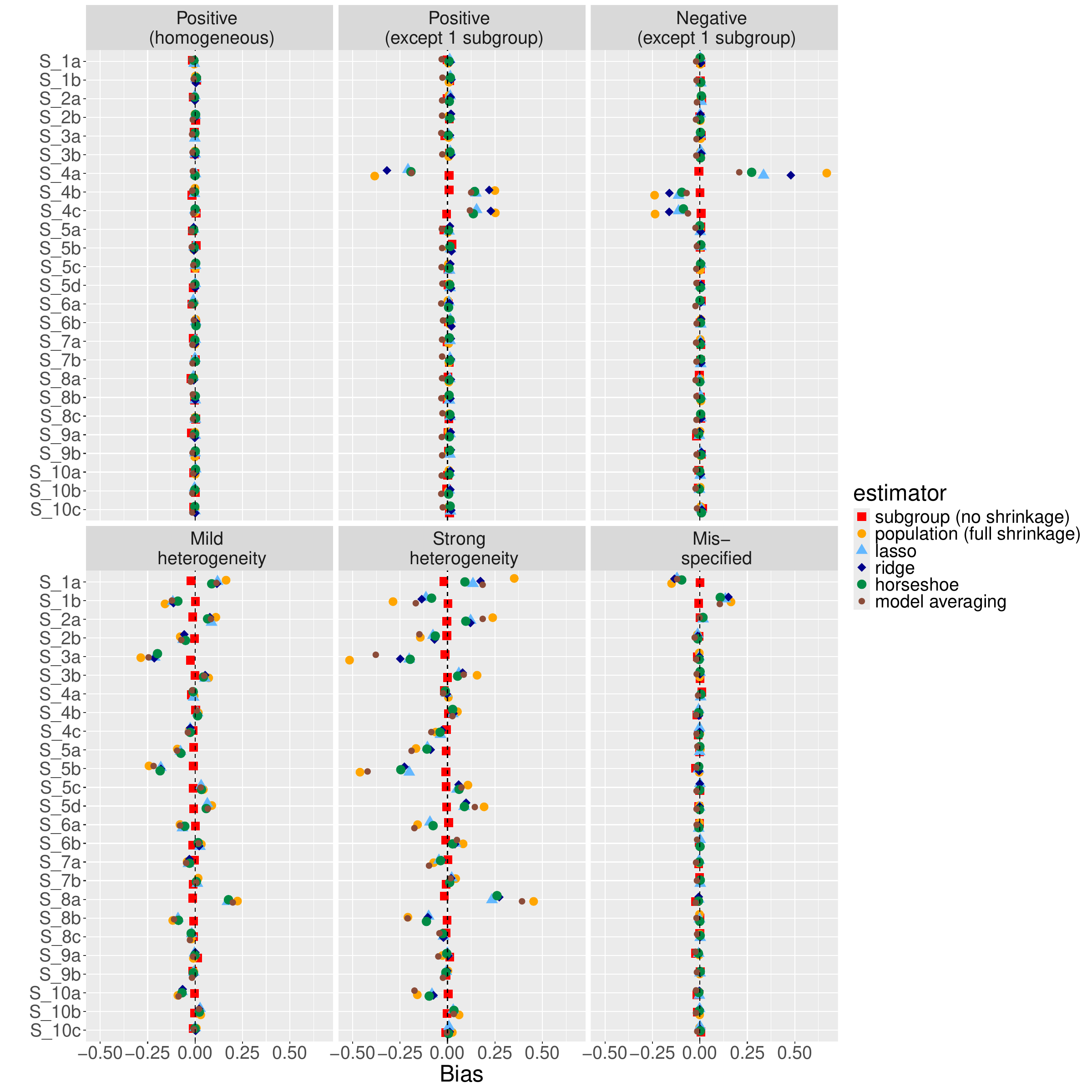}
\caption{\label{fig:unnamed-chunk-4}\label{fig:bias}Bias for all simulation scenarios and subgroups.}
\end{figure}

\begin{figure}
\centering
\includegraphics{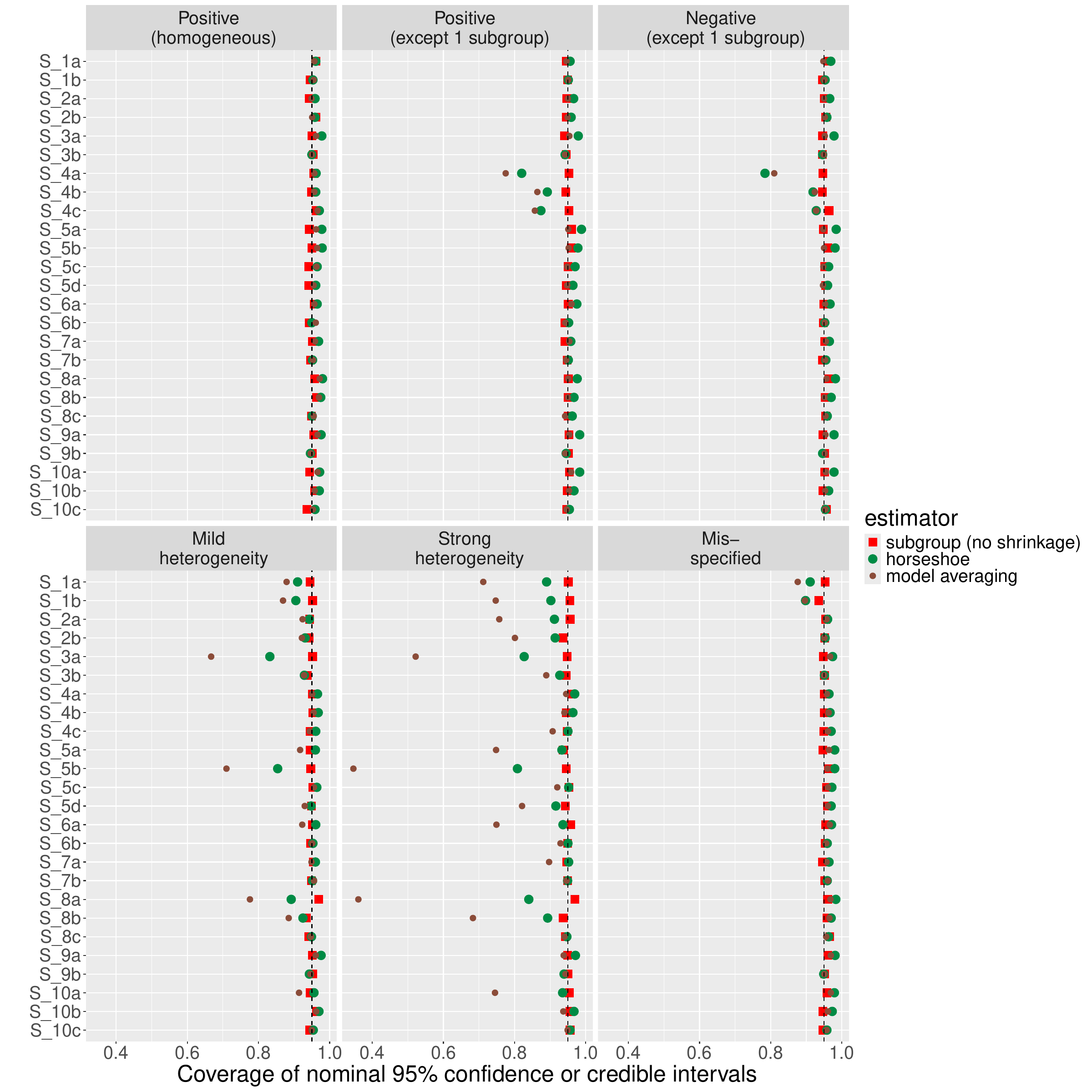}
\caption{\label{fig:unnamed-chunk-5}\label{fig:coverage}Frequentist coverage of nominal 95\% confidence or credible intervals for all simulation scenarios and subgroups.}
\end{figure}

\begin{figure}
\centering
\includegraphics{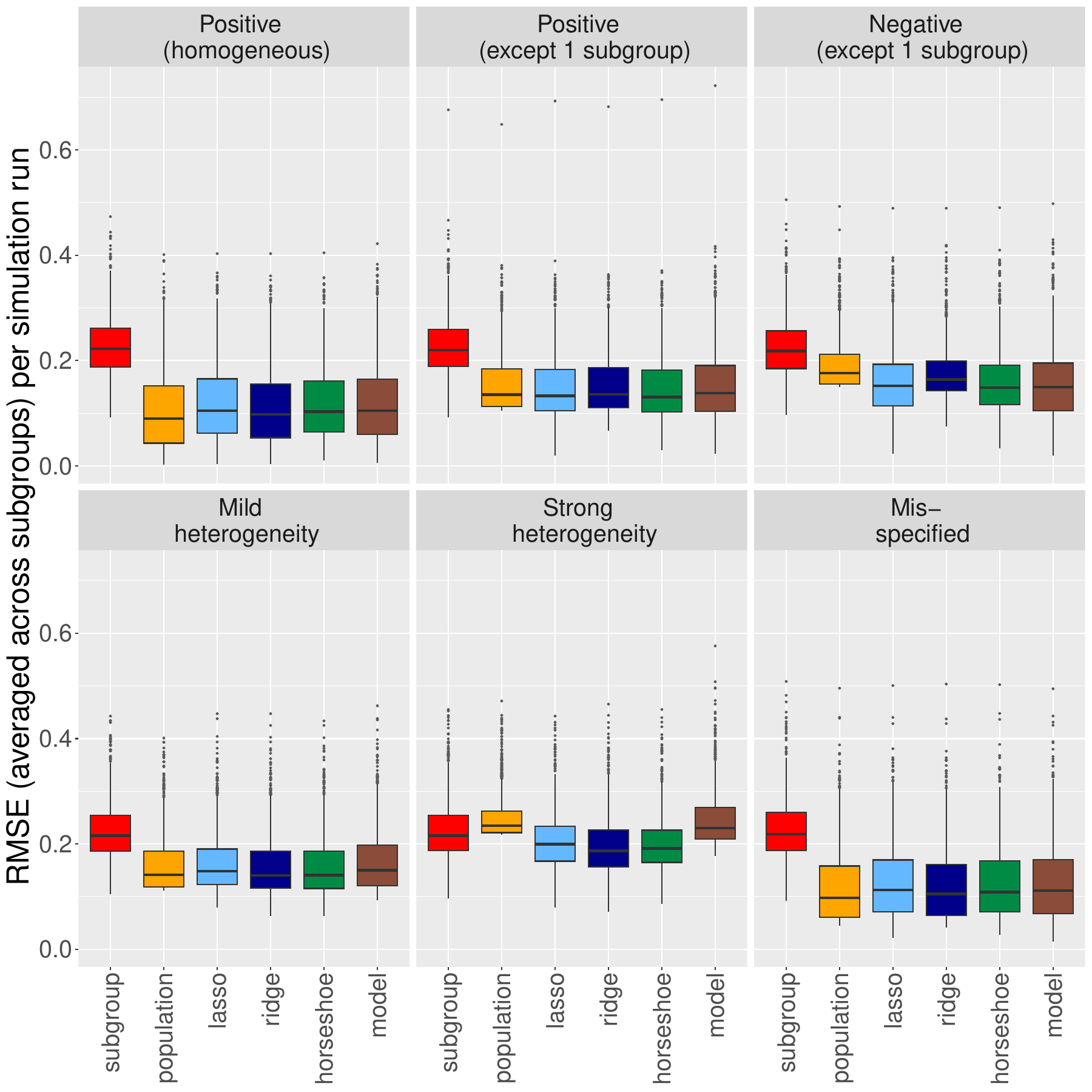}
\caption{\label{fig:unnamed-chunk-6}\label{fig:boxRMSE}Boxplots of root mean squared errors (averaged across subgroups) for each simulation run stratified by estimator and simulation scenario.}
\end{figure}

\begin{figure}
\centering
\includegraphics{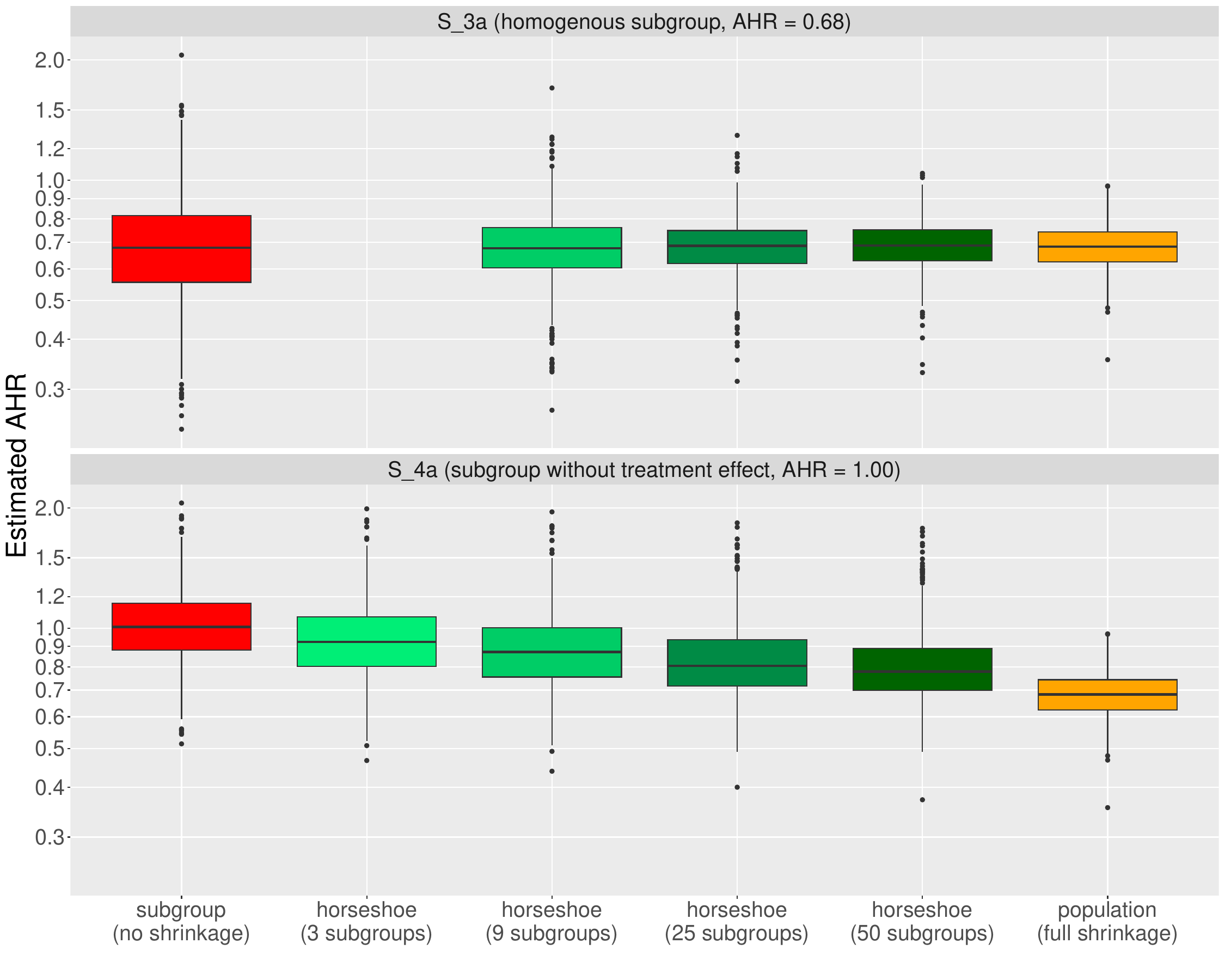}
\caption{\label{fig:unnamed-chunk-7}\label{fig:horseshoeMoreSubgroups}Boxplots of treatment effect estimators for Scenario 2 in subgroups \(S_{3a}\) (a homogeneous subgroup) and \(S_{4a}\) (the only subgroup without a treatment effect) across 1'000 simulation runs. Shown are the subgroup estimator without shrinkage, the population estimator, and different horseshoe estimators with a varying number of subgrouping variables included in the global model. The horseshoe estimator with 3 subgroups is not available for subgroup \(S_{3a}\) because it was not included in the respective global model.}
\end{figure}

\renewcommand\refname{References}


\begin{thebibliography}{37}
\providecommand{\natexlab}[1]{#1}
\providecommand{\url}[1]{\texttt{#1}}
\expandafter\ifx\csname urlstyle\endcsname\relax
  \providecommand{\doi}[1]{doi: #1}\else
  \providecommand{\doi}{doi: \begingroup \urlstyle{rm}\Url}\fi

\bibitem[Agresti(2002)]{agresti2012}
Alan Agresti.
\newblock \emph{Categorical data analysis - second edition}.
\newblock John Wiley \& Sons, 2002.

\bibitem[Alosh et~al.(2017)Alosh, Huque, Bretz, and
  D'Agostino~Sr]{alosh2017tutorial}
Mohamed Alosh, Mohammad~F Huque, Frank Bretz, and Ralph~B D'Agostino~Sr.
\newblock Tutorial on statistical considerations on subgroup analysis in
  confirmatory clinical trials.
\newblock \emph{Statistics in Medicine}, 36\penalty0 (8):\penalty0 1334--1360,
  2017.

\bibitem[Ballarini et~al.(2021)Ballarini, Thomas, Rosenkranz, and
  Bornkamp]{ballarini2021}
Nicolas~M Ballarini, Marius Thomas, Gerd~K Rosenkranz, and Bj{\"o}rn Bornkamp.
\newblock subtee: an r package for subgroup treatment effect estimation in
  clinical trials.
\newblock \emph{Journal of Statistical Software}, 99:\penalty0 1--17, 2021.

\bibitem[Bornkamp et~al.(2017)Bornkamp, Ohlssen, Magnusson, and
  Schmidli]{bornkamp2017}
Bj{\"o}rn Bornkamp, David Ohlssen, Baldur~P Magnusson, and Heinz Schmidli.
\newblock Model averaging for treatment effect estimation in subgroups.
\newblock \emph{Pharmaceutical Statistics}, 16\penalty0 (2):\penalty0 133--142,
  2017.

\bibitem[Brilleman et~al.(2020)Brilleman, Elci, Novik, and
  Wolfe]{brilleman2020}
Samuel~L Brilleman, Eren~M Elci, Jacqueline~Buros Novik, and Rory Wolfe.
\newblock Bayesian survival analysis using the rstanarm r package.
\newblock \emph{arXiv preprint arXiv:2002.09633}, 2020.

\bibitem[B{\"u}rkner(2017)]{bruckner2017}
Paul-Christian B{\"u}rkner.
\newblock {brms}: An {R} package for {Bayesian} multilevel models using {Stan}.
\newblock \emph{Journal of Statistical Software}, 80\penalty0 (1):\penalty0
  1--28, 2017.

\bibitem[Carvalho et~al.(2009)Carvalho, Polson, and Scott]{carvalho2009}
Carlos~M Carvalho, Nicholas~G Polson, and James~G Scott.
\newblock Handling sparsity via the horseshoe.
\newblock In \emph{Proceedings of the Twelth International Conference on
  Artificial Intelligence and Statistics}, volume~5, pages 73--80, 2009.

\bibitem[Daniel et~al.(2021)Daniel, Zhang, and Farewell]{daniel2021}
Rhian Daniel, Jingjing Zhang, and Daniel Farewell.
\newblock Making apples from oranges: Comparing noncollapsible effect
  estimators and their standard errors after adjustment for different covariate
  sets.
\newblock \emph{Biometrical Journal}, 63\penalty0 (3):\penalty0 528--557, 2021.

\bibitem[Didelez and Stensrud(2022)]{didelez2022}
Vanessa Didelez and Mats~Julius Stensrud.
\newblock On the logic of collapsibility for causal effect measures.
\newblock \emph{Biometrical Journal}, 64\penalty0 (2):\penalty0 235--242, 2022.

\bibitem[{EMA Biostatistics Working Party}(2019)]{emaSubgroups2019}
{EMA Biostatistics Working Party}.
\newblock Guideline on the investigation of subgroups in confirmatory clinical
  trials., 2019.
\newblock URL
  \url{https://www.ema.europa.eu/en/documents/scientific-guideline/guideline-investigation-subgroups-confirmatory-clinical-trials_en.pdf}.

\bibitem[Freidlin et~al.(2010)Freidlin, McShane, and Korn]{freidlin2010}
Boris Freidlin, Lisa~M McShane, and Edward~L Korn.
\newblock Randomized clinical trials with biomarkers: design issues.
\newblock \emph{Journal of the National Cancer Institute}, 102\penalty0
  (3):\penalty0 152--160, 2010.

\bibitem[Gregson et~al.(2019)Gregson, Sharples, Stone, Burman, {\"O}hrn, and
  Pocock]{gregson2019}
John Gregson, Linda Sharples, Gregg~W Stone, Carl-Fredrik Burman, Fredrik
  {\"O}hrn, and Stuart Pocock.
\newblock Nonproportional hazards for time-to-event outcomes in clinical
  trials: Jacc review topic of the week.
\newblock \emph{Journal of the American College of Cardiology}, 74\penalty0
  (16):\penalty0 2102--2112, 2019.

\bibitem[Henderson et~al.(2016)Henderson, Louis, Wang, and
  Varadhan]{henderson2016}
Nicholas~C Henderson, Thomas~A Louis, Chenguang Wang, and Ravi Varadhan.
\newblock Bayesian analysis of heterogeneous treatment effects for
  patient-centered outcomes research.
\newblock \emph{Health Services and Outcomes Research Methodology},
  16:\penalty0 213--233, 2016.

\bibitem[Hern{\'a}n and Robins(2023)]{hernan2023}
Miguel~A Hern{\'a}n and James~M Robins.
\newblock \emph{Causal Inference: What If}.
\newblock CRC Press Inc, Boca Raton, FL, 2023.

\bibitem[{ICH E17 Expert Working Group}(2016)]{iche17}
{ICH E17 Expert Working Group}.
\newblock {ICH E17: General principles for planning and designing
  multi-regional clinical trials.}, 2016.
\newblock URL
  \url{https://database.ich.org/sites/default/files/E17EWG_Step4_2017_1116.pdf}.

\bibitem[Jones et~al.(2011)Jones, Ohlssen, Neuenschwander, Racine, and
  Branson]{jones2011}
Hayley~E Jones, David~I Ohlssen, Beat Neuenschwander, Amy Racine, and Michael
  Branson.
\newblock Bayesian models for subgroup analysis in clinical trials.
\newblock \emph{Clinical Trials}, 8\penalty0 (2):\penalty0 129--143, 2011.

\bibitem[Kalbfleisch and Prentice(1981)]{kalbfleisch1981}
John~D Kalbfleisch and Ross~L Prentice.
\newblock Estimation of the average hazard ratio.
\newblock \emph{Biometrika}, 68\penalty0 (1):\penalty0 105--112, 1981.

\bibitem[Lipkovich et~al.(2017)Lipkovich, Dmitrienko, and
  B~D'Agostino~Sr]{lipkovich2017tutorial}
Ilya Lipkovich, Alex Dmitrienko, and Ralph B~D'Agostino~Sr.
\newblock Tutorial in biostatistics: data-driven subgroup identification and
  analysis in clinical trials.
\newblock \emph{Statistics in Medicine}, 36\penalty0 (1):\penalty0 136--196,
  2017.

\bibitem[Liu et~al.(2022)Liu, Wang, Yang, Hui, Xu, Kil, and Hsu]{liu2022}
Yi~Liu, Bushi Wang, Miao Yang, Jianan Hui, Heng Xu, Siyoen Kil, and Jason~C
  Hsu.
\newblock Correct and logical causal inference for binary and time-to-event
  outcomes in randomized controlled trials.
\newblock \emph{Biometrical Journal}, 64\penalty0 (2):\penalty0 198--224, 2022.

\bibitem[Loh et~al.(2019)Loh, Cao, and Zhou]{loh2019}
Wei-Yin Loh, Luxi Cao, and Peigen Zhou.
\newblock Subgroup identification for precision medicine: A comparative review
  of 13 methods.
\newblock \emph{Wiley Interdisciplinary Reviews: Data Mining and Knowledge
  Discovery}, 9\penalty0 (5):\penalty0 e1326, 2019.

\bibitem[Marcus et~al.(2017)Marcus, Davies, Ando, Klapper, Opat, Owen,
  Phillips, Sangha, Schlag, Seymour, et~al.]{marcus2017}
Robert Marcus, Andrew Davies, Kiyoshi Ando, Wolfram Klapper, Stephen Opat,
  Carolyn Owen, Elizabeth Phillips, Randeep Sangha, Rudolf Schlag, John~F
  Seymour, et~al.
\newblock Obinutuzumab for the first-line treatment of follicular lymphoma.
\newblock \emph{New England Journal of Medicine}, 377\penalty0 (14):\penalty0
  1331--1344, 2017.

\bibitem[Nguyen~Duc et~al.(2021)Nguyen~Duc, Heinzmann, Berge, and
  Wolbers]{nguyen2021}
Anh Nguyen~Duc, Dominik Heinzmann, Claude Berge, and Marcel Wolbers.
\newblock A pragmatic adaptive enrichment design for selecting the right target
  population for cancer immunotherapies.
\newblock \emph{Pharmaceutical Statistics}, 20\penalty0 (2):\penalty0 202--211,
  2021.

\bibitem[Pennello and Rothmann(2018)]{pennello2018}
Gene Pennello and Mark Rothmann.
\newblock Bayesian subgroup analysis with hierarchical models.
\newblock In \emph{Chapter 10 of Biopharmaceutical Applied Statistics
  Symposium: Volume 2 Biostatistical Analysis of Clinical Trials (edited by
  {KE} Peace, {DG} Chen, and {S} Menon)}, pages 175--192. Springer, 2018.

\bibitem[Piironen and Vehtari(2017)]{piironen2017}
Juho Piironen and Aki Vehtari.
\newblock Sparsity information and regularization in the horseshoe and other
  shrinkage priors.
\newblock \emph{Electronic Journal of Statistics}, 11\penalty0 (2):\penalty0
  5018--5051, 2017.

\bibitem[Qi et~al.(2022)Qi, Zhou, Wang, and Peterson]{qi2022}
Xinyue Qi, Shouhao Zhou, Yucai Wang, and Christine Peterson.
\newblock Bayesian sparse modeling to identify high-risk subgroups in
  meta-analysis of safety data.
\newblock \emph{Research Synthesis Methods}, 13\penalty0 (6):\penalty0
  807--820, 2022.

\bibitem[Rauch et~al.(2018)Rauch, Brannath, Br{\"u}ckner, and
  Kieser]{rauch2018}
Geraldine Rauch, Werner Brannath, Matthias Br{\"u}ckner, and Meinhard Kieser.
\newblock The average hazard ratio--a good effect measure for time-to-event
  endpoints when the proportional hazard assumption is violated?
\newblock \emph{Methods of Information in Medicine}, 57\penalty0 (03):\penalty0
  089--100, 2018.

\bibitem[Schandelmaier et~al.(2020)Schandelmaier, Briel, Varadhan, Schmid,
  Devasenapathy, Hayward, Gagnier, Borenstein, van~der Heijden, Dahabreh,
  et~al.]{schandelmaier2020iceman}
Stefan Schandelmaier, Matthias Briel, Ravi Varadhan, Christopher~H Schmid,
  Niveditha Devasenapathy, Rodney~A Hayward, Joel Gagnier, Michael Borenstein,
  Geert~JMG van~der Heijden, Issa~J Dahabreh, et~al.
\newblock Development of the instrument to assess the credibility of effect
  modification analyses (iceman) in randomized controlled trials and
  meta-analyses.
\newblock \emph{Canadian Medical Association Journal}, 192\penalty0
  (32):\penalty0 E901--E906, 2020.

\bibitem[Schemper et~al.(2009)Schemper, Wakounig, and Heinze]{schemper2009}
Michael Schemper, Samo Wakounig, and Georg Heinze.
\newblock The estimation of average hazard ratios by weighted cox regression.
\newblock \emph{Statistics in Medicine}, 28\penalty0 (19):\penalty0 2473--2489,
  2009.

\bibitem[Simon et~al.(2011)Simon, Friedman, Hastie, and
  Tibshirani]{SimonGlmnetCox2011}
Noah Simon, Jerome Friedman, Trevor Hastie, and Rob Tibshirani.
\newblock Regularization paths for cox's proportional hazards model via
  coordinate descent.
\newblock \emph{Journal of Statistical Software}, 39\penalty0 (5):\penalty0
  1--13, 2011.

\bibitem[Sleight(2000)]{sleight2000debate}
Peter Sleight.
\newblock Debate: Subgroup analyses in clinical trials: fun to look at-but
  don't believe them!
\newblock \emph{Trials}, 1\penalty0 (1):\penalty0 1--3, 2000.

\bibitem[Sun et~al.(2014)Sun, Ioannidis, Agoritsas, Alba, and
  Guyatt]{sun2014use}
Xin Sun, John~PA Ioannidis, Thomas Agoritsas, Ana~C Alba, and Gordon Guyatt.
\newblock How to use a subgroup analysis: users` guide to the medical
  literature.
\newblock \emph{Journal of the American Medical Association}, 311\penalty0
  (4):\penalty0 405--411, 2014.

\bibitem[Taylor and Tibshirani(2018)]{taylor2018}
Jonathan Taylor and Robert Tibshirani.
\newblock Post-selection inference for-penalized likelihood models.
\newblock \emph{Canadian Journal of Statistics}, 46\penalty0 (1):\penalty0
  41--61, 2018.

\bibitem[Ternes et~al.(2017)Ternes, Rotolo, Heinze, and
  Michiels]{ternes2017identification}
Nils Ternes, Federico Rotolo, Georg Heinze, and Stefan Michiels.
\newblock Identification of biomarker-by-treatment interactions in randomized
  clinical trials with survival outcomes and high-dimensional spaces.
\newblock \emph{Biometrical Journal}, 59\penalty0 (4):\penalty0 685--701, 2017.

\bibitem[Townsend et~al.(2023)Townsend, Hiddemann, Buske, Cartron, Cunningham,
  Dyer, Gribben, Phillips, Dreyling, Seymour, et~al.]{townsend2023}
William Townsend, Wolfgang Hiddemann, Christian Buske, Guillaume Cartron, David
  Cunningham, Martin~JS Dyer, John~G Gribben, Elizabeth~H Phillips, Martin
  Dreyling, John~F Seymour, et~al.
\newblock Obinutuzumab versus rituximab immunochemotherapy in previously
  untreated inhl: final results from the gallium study.
\newblock \emph{Hemasphere}, 7\penalty0 (7):\penalty0 e919, 2023.

\bibitem[Van~Erp et~al.(2019)Van~Erp, Oberski, and Mulder]{vanErp2019}
Sara Van~Erp, Daniel~L Oberski, and Joris Mulder.
\newblock Shrinkage priors for bayesian penalized regression.
\newblock \emph{Journal of Mathematical Psychology}, 89:\penalty0 31--50, 2019.

\bibitem[{Vázquez Rabuñal} et~al.(2024){Vázquez Rabuñal}, {Sabanés Bové},
  and Wolbers]{bonsaiforest}
Mar {Vázquez Rabuñal}, Daniel {Sabanés Bové}, and Marcel Wolbers.
\newblock \emph{bonsaiforest: Shrinkage Based Forest Plots}, 2024.
\newblock URL \url{https://CRAN.R-project.org/package=bonsaiforest}.
\newblock R package version 0.1.0.

\bibitem[Yusuf et~al.(1991)Yusuf, Wittes, Probstfield, and
  Tyroler]{yusuf1991jama}
Salim Yusuf, Janet Wittes, Jeffrey Probstfield, and Herman~A Tyroler.
\newblock Analysis and interpretation of treatment effects in subgroups of
  patients in randomized clinical trials.
\newblock \emph{Journal of the American Medical Association}, 266\penalty0
  (1):\penalty0 93--98, 1991.

\end{thebibliography}
\end{document}